\def\year{2023}\relax
\documentclass[letterpaper]{article} 
\usepackage{aaai23}  
\usepackage{times}  
\usepackage{helvet}  
\usepackage{courier}  
\usepackage[hyphens]{url}  
\usepackage{graphicx} 
\usepackage{natbib}  
\usepackage{caption} 
\DeclareCaptionStyle{ruled}{labelfont=normalfont,labelsep=colon,strut=off} 
\usepackage{algorithm}
\usepackage{algorithmic}

\usepackage{xcolor}

\usepackage{newfloat}
\usepackage{listings}
\floatstyle{ruled}
\newfloat{listing}{tb}{lst}{}
\floatname{listing}{Listing}
\nocopyright
\usepackage{makecell}
\usepackage{bm}
\usepackage{enumitem}
\usepackage{multirow}
\usepackage{tabularx}
\usepackage{amssymb}
\usepackage{amsmath}
\usepackage{pifont}
\usepackage{booktabs}
\usepackage{graphicx}
\usepackage{nameref}
\usepackage{lineno}
\usepackage[subrefformat=parens,skip=0pt]{subcaption}
\usepackage[detect-all]{siunitx}
\usepackage{xspace}
\newcommand\ie{i.\,e.\xspace}
\newcommand\eg{e.\,g.\xspace}

\newcommand\US{U.S.\xspace}

\newcommand{\var}[1]{\mathit{#1}}

\def\sym#1{\ifmmode^{#1}\else\(^{#1}\)\fi}

\let\namerefOld\nameref
\renewcommand{\nameref}[1]{\textit{\namerefOld{#1}}}

\title{Is Fact-Checking Politically Neutral? Asymmetries in How U.S. Fact-Checking Organizations Pick Up False Statements Mentioning Political Elites}
\author{
    Yuwei Chuai\textsuperscript{\rm 1}, Jichang Zhao\textsuperscript{\rm 2}, Nicolas Pr{\"o}llochs\textsuperscript{\rm 3}, Gabriele Lenzini\textsuperscript{\rm 1}\\}
\affiliations{
    \textsuperscript{\rm 1}University of Luxembourg, Luxembourg\\
    \textsuperscript{\rm 2}Beihang University, China \\
    \textsuperscript{\rm 3}JLU Giessen, Germany \\
    yuwei.chuai@uni.lu, jichang@buaa.edu.cn, nicolas.proellochs@wi.jlug.de, gabriele.lenzini@uni.lu
}

\begin{document}

\maketitle

\begin{abstract}
Political elites play an important role in the proliferation of online misinformation. However, an understanding of how fact-checking platforms pick up politicized misinformation for fact-checking is still in its infancy. Here, we conduct an empirical analysis of mentions of \US political elites within fact-checked statements. For this purpose, we collect a comprehensive dataset consisting of \num{35014} true and false statements that have been fact-checked by two major fact-checking organizations (Snopes, PolitiFact) in the \US between 2008 and 2023, \ie, within an observation period of 15 years. Subsequently, we perform content analysis and explanatory regression modeling to analyze how veracity is linked to mentions of \US political elites in fact-checked statements. Our analysis yields the following main findings: (i) Fact-checked false statements are, on average, 20\% more likely to mention political elites than true fact-checked statements. (ii) There is a partisan asymmetry such that fact-checked false statements are 88.1\% more likely to mention Democrats, but 26.5\% less likely to mention Republicans, compared to fact-checked true statements. (iii) Mentions of political elites in fact-checked false statements reach the highest level during the months preceding elections. (iv) Fact-checked false statements that mention political elites carry stronger other-condemning emotions and are more likely to be pro-Republican, compared to fact-checked true statements. In sum, our study offers new insights into understanding mentions of political elites in false statements on \US fact-checking platforms, and bridges important findings at the intersection between misinformation and politicization.
\end{abstract}

\section{Introduction}
Online misinformation poses significant challenges to societies of the twenty-first century \cite{lazer_science_2018}. Although the proportion of misinformation in (most) people's information diet tends to be small \cite{budak_misunderstanding_2024,ecker_misinformation_2024,altay_misinformation_2023}, detrimental effects of misinformation have been repeatedly observed across various events such as, for example, during elections and crisis situations \cite{allcott_social_2017,suciu_bots_2022,mosleh_measuring_2022,hartman_interventions_2022,pierri_online_2022,bar_new_2023}. Hence, countering the spread of misinformation has become an urgent priority, with extensive research efforts dedicated to analysis of its online diffusion and detection \cite{raponi_fake_2022,ecker_psychological_2022,horta_ribeiro_automated_2023,vosoughi_spread_2018,drolsbach_diffusion_2023,prollochs_emotions_2021,chuai_anger_2022,chuai_roll-out_2023}. However, despite these efforts, reducing people's susceptibility to misinformation remains challenging \cite{ecker_psychological_2022}. 

Recent research suggests that political elites (\ie, politicians, political figures) may play a key role in the proliferation of online misinformation \cite{flores_politicians_2022,mosleh_measuring_2022}. Over the past 30 years, political figures have been increasingly mentioned in media representations with the goal of politicizing social issues \cite{chinn_politicization_2020}. The strategy of mentioning politicians in news stories can drive engagement \cite{osmundsen_partisan_2021,shin_partisan_2017} and help to foster specific political interests, ideologies, or agendas \cite{druckman_threats_2022,chinn_politicization_2020, hart_politicization_2020}. At the same time, politicization contributes to partisan and affective polarization, characterized by growing disparities in viewpoints between adherents of opposing political parties \cite{simon_politicization_2019,raymond_measuring_2022,saveski_engaging_2022}. For instance, misinformation mentioning politicians has been shown to fragment online political debates and intensify divisions in public support for policies related to COVID-19 and climate change \cite{flores_politicians_2022,heiberger_not_2022,chinn_politicization_2020}. In light of a substantial body of research highlighting the influence of political elites on individuals' attitudes and behaviors \cite{pierri_online_2022,mosleh_measuring_2022,enders_relationship_2022}, the politicization of misinformation has become a pressing concern that requires further scrutiny.

Fact-checking organizations such as PolitiFact (politifact.com) and Snopes (snopes.com) are vital for identifying and debunking politicized misinformation. Their fact-checking assessments are supposed to help users to identify misleading statements in order to curb their spread (\eg, on social media) \cite{shao_hoaxy_2016}. Previous studies suggest that fact-checking can indeed improve misinformation discernment and reduce sharing intentions for misinformation \cite{altay_misinformation_2023,martel_misinformation_2023}. In practice, however, limited resources often force fact-checkers to prioritize statements for fact-checking \cite{sehat_misinformation_2024}. Although their assessments are generally considered accurate \cite{vosoughi_spread_2018,lee_fact_2023}, the necessity of choosing specific claims to fact-check can invite criticism due to the subjective nature of the decision \cite{nieminen_fighting_2019}. Critics argue that the selection of claims can reflect political preferences, potentially skewing results in favor of certain political identities (typically towards the political left) \cite{singh_independent_2024}. This is problematic as the perception of bias and subjectivity can erode public trust in fact-checking organizations. If people believe that fact-checkers are not impartial, they may be less likely to trust their findings, regardless of the evidence presented \cite{drolsbach_community_2024}. Despite these vocal concerns, however, empirical evidence on how fact-checking organizations pick up politicized misinformation for fact-checking is missing.

\textbf{Research goal:} 
In this paper, we conduct an empirical analysis of mentions of political elites within fact-checked true and false statements. To this end, we collect \num{35014} true and false statements that have been fact-checked by major \US fact-checking organizations, namely, PolitiFact (politifact.com) and Snopes (snopes.com), between 2008 and 2023, \ie, within an observation period of 15 years. These two fact-checking organizations are among the largest and most established of their kind \cite{allcott_social_2017,vosoughi_spread_2018,mosleh_measuring_2022,garrett_conservatives_2021}. Additionally, we construct a database of \US politicians, containing \num{3703} political elites (Republicans and Democrats) to measure politicization within the fact-checked statements. Subsequently, we estimate explanatory regression models to analyze how veracity is linked to mentions of \US political elites in fact-checked statements. We also study how this link varies between election cycles and across topics, and perform content analysis to explore the role of other-condemning emotions in the politicization of misinformation.

\textbf{Contributions:} Our work provides empirical evidence on how fact-checking organizations pick up politicized misinformation for fact-checking. We find that: (i) fact-checked false statements are, on average, 20\% more likely to mention political elites than true fact-checked statements. (ii) There is a partisan asymmetry such that false fact-checked statements are 88.1\% more likely to mention Democrats, but 26.5\% less likely to mention Republicans, compared to true fact-checked statements. (iii) Mentions of political elites in fact-checked false statements reach the highest level during the months preceding elections. (iv) Fact-checked false statements that mention political elites carry stronger other-condemning emotions and are more likely to be pro-Republican. These findings offer new insights into how major fact-checking platforms in the \US select their targets to fact-check, and bridge earlier work at the intersection between misinformation and politicization.

\section{Background}

Online misinformation is a pressing societal problem that platforms, policymakers, and researchers continue to grapple with. The reason is that there are serious concerns that misinformation on social media is damaging societies and democratic institutions \cite{lazer_science_2018,ecker_misinformation_2024,feuerriegel_research_2023}. While some critics argue that the harms of misinformation may be overstated, with exposure to falsehoods often concentrated among narrow fringe groups \cite{cinelli_echo_2021,altay_misinformation_2023,budak_misunderstanding_2024}, negative repercussions of online misinformation on modern societies have been repeatedly observed in recent years, especially in the context of elections and crises \cite{suciu_bots_2022,allcott_social_2017,mosleh_measuring_2022,hartman_interventions_2022,pierri_online_2022}. Therefore, the fight against misinformation is both warranted and required \cite{ecker_misinformation_2024}.

The intersection of misinformation and political elites has garnered increased attention lately \cite{mosleh_measuring_2022,flores_politicians_2022,osmundsen_partisan_2021}. There has been a surge of news stories featuring political figures to politicize social issues. Politicization can drive engagement and help to foster political agendas \cite{druckman_threats_2022,chinn_politicization_2020, hart_politicization_2020,shin_partisan_2017,osmundsen_partisan_2021}. Understanding politicization is important as it can contribute to partisan and affective polarization, which, in turn, strongly affects how people evaluate information \cite{chinn_politicization_2020,shin_partisan_2017,flores_politicians_2022,jenke_affective_2023}. Existing works show that politicized misinformation can have profound impacts on voter attitudes and other offline behaviors\cite{chinn_politicization_2020,hart_politicization_2020,shin_partisan_2017,pierri_online_2022, mosleh_measuring_2022, enders_relationship_2022}. For instance, the politicization of the (seemingly unpolitical) topic of climate change has contributed to a polarized environment that fosters climate change denial and misinformation \cite{chinn_politicization_2020}. Likewise, during elections, partisans have been observed to selectively share (oftentimes false) statements that express animosity toward political opponents (\ie, members with the opposite
political identity) \cite{shin_partisan_2017}. Overall, the growing politicization is a pressing concern that has the potential to amplify the negative effects of misinformation on modern societies.

\section{Research Questions}

Previous studies indicate that political elites play a key role in the proliferation of misinformation. Yet, an understanding of how fact-checking platforms pick up politicized misinformation for fact-checking is still in its infancy. To address this gap, we propose four research questions (RQs1--4), each of which aims to extract new insights at the intersection of fact-checking and political elites.

\textbf{Politicization (RQ1):} 
Fact-checks carried out by fact-checking organizations are supposed to identify misinformation in order to inform the public and help curb its spread. However, the manual fact-checking process is both labor-intensive and time-consuming, and fact-checking organizations operate with limited resources \cite{micallef_true_2022}. This inherent limitation constrains the scale of fact-checking efforts and forces fact-checkers to prioritize statements based on their potential harm and urgency. Fragmentation and likelihood of spread are two main criteria for fact-checking prioritization \cite{sehat_misinformation_2024}. Politicized misinformation is particularly prone to go viral, incites partisan and affective polarization, and poses significant threats to societies \cite{flores_politicians_2022,mosleh_measuring_2022,osmundsen_partisan_2021}. Here, mentioning political elites is a primary strategy to politicize misinformation stories \cite{hart_politicization_2020,chinn_politicization_2020,druckman_threats_2022}. 
Given this, fact-checking platforms may prioritize politicized statements mentioning political elites for fact-checking. However, empirical evidence on how the mentioning of political elites differs between fact-checked true and false stories is missing. Accordingly, our first research question (RQ1) states:

\noindent \textit{\textbf{RQ1}: Are fact-checked false statements more likely to mention political elites than fact-checked true statements?}
\vspace{.5em}

\textbf{Partisan asymmetry (RQ2):} 
Politicization typically has a distinct partisan leaning, favoring either conservative (Republican, right-leaning) or liberal (Democratic, left-leaning) opinions \cite{garrett_conservatives_2021}. Two factors lead us to anticipate that there may be a partisan asymmetry regarding true and false statements picked up for fact-checking by professional fact-checking organizations. First, there is ample evidence that the political left and right use the internet differently and right-leaning sources gain more visibility on social media, a phenomenon known as \emph{ideological asymmetry} \citep{gonzalez-bailon_advantage_2022,grossmann_asymmetric_2016}. Further, adherents of the right have been observed to be more likely to believe and share misinformation than adherents of the left \cite{garrett_conservatives_2021,mosleh_measuring_2022,nikolov_right_2021,gonzalez-bailon_advantage_2022,shin_partisan_2017,tai_official_2023}. Second, fact-checking organizations may more frequently target one political side or another. Anecdotal and small-scale empirical evidence suggest that fact-checkers in the \US target misinformation from Republicans more often than misinformation from Democrats \cite{ostermeier_selection_2011}. Motivated by these two factors, our second research question (RQ2) is to gauge whether mentions of Democrats and Republicans vary across fact-checked true and false statements.

\noindent \textit{\textbf{RQ2}: Is there a partisan asymmetry in mentions of political elites across fact-checked true and false statements?}
\vspace{.5em}

\textbf{Elections (RQ3):} 
Elections are important political events in democracies, serving as focal points for political engagement and discourse. The campaigning activities of political candidates inherently generate a large number of (true and false) statements, which can affect voters' attitudes and behaviors \cite{mathur_how_2022, suciu_bots_2022}.
In response to the increased volume, professional fact-checking organizations may allocate more time and resources than usual to scrutinize politicized statements \cite{lee_fact_2023,sehat_misinformation_2024}. Consequently, the level of politicization in fact-checked statements may rise as elections approach. Therefore, our third research question (RQ3) is:

\noindent \textit{\textbf{RQ3}: Are political elites more likely to be mentioned in fact-checked statements as elections draw closer?}
\vspace{.5em}

\textbf{Other-condemning emotions (RQ4):} 
A frequently employed strategy to politicize misinformation is to harness animosity toward political elites of the political out-group \cite{hartman_interventions_2022,rathje_out-group_2021}. Out-group animosity is an effective strategy for expressing one's partisan identity and generating engaging online content. For instance, it has been found that out-party animosity exerts a substantial influence on the spread of political misinformation when contrasted with in-party warmth \cite{rathje_out-group_2021}. Animosity toward the out-group is frequently expressed through the use of other-condemning emotions (\ie, contempt, anger, and disgust), a subset of the moral emotions families \cite{solovev_moral_2022,dastani_other-condemning_2017}. Other-condemning emotions are reactions to the social behavior of others, involving a negative judgement or disapproval \cite{solovev_moral_2022,dastani_other-condemning_2017,rozin_cad_1999}. Studying the role of other-condemning emotions thus allows one to better understand the context of politicization, \ie, whether authors of fact-checked statements harness animosity toward the political out-groups. Accordingly, our fourth research question (RQ4) analyzes whether fact-checked statements that mention political elites contain more other-condemning emotions compared to fact-checked statements that do not mention political elites.

\noindent \textit{\textbf{RQ4}: Do fact-checked statements that mention political elites contain more other-condemning emotions compared to fact-checked statements that do not mention political elites?}
\vspace{.5em}

\section{Data and Methods}
\label{sec:materials_and_methods}

\begin{table*}
\footnotesize
\begin{center}
\begin{tabularx}{\textwidth}{@{\hspace{\tabcolsep}\extracolsep{\fill}}l l l l l}
\toprule
Date&Statement&Affiliated Party&Other-Condemning&Veracity\\
\midrule
2022-02-25&Gov. \textbf{Tony Evers} has ``only gotten one-third of the money&Democratic&0.069&True\\ 
&meant for COVID relief out the door. He is sitting on&&&\\ 
&\$930 million in ARPA funds left unspent. In fact, he still has&&&\\ 
&CARES Act money from two years ago.''&&&\\
2018-06-25&\$1 billion—that’s how much \textbf{Bruce Rauner} has wasted&Republican&0.750&False\\
&with his budget crisis.&&&\\
2016-10-25&Wikileaks also shows how \textbf{John Podesta} rigged the polls &Democratic&0.667&False\\ 
&by oversampling Democrats, a voter suppression technique.&&&\\
2014-10-23&Six convictions are connected to an allegation by prosecutors&Republican&0.571&False\\
&``that Gov. \textbf{Scott Walker} is at the center of a criminal scheme.''&&&\\
\bottomrule
\end{tabularx}
\caption{Examples of fact-checked true and false statements. Mentions of political elites are highlighted in bold in the statements. The affiliated parties of mentioned politicians and other-condemning emotions in the statements are included.}
\label{tab:claim_examples}
\end{center}
\end{table*}

\subsection{Data Source: Fact-Checked Statements}
For our analysis, we collect a large-scale dataset consisting of true and false statements that have been fact-checked between 2008 and 2023 (\ie, within an observation period of 15 years) by two major fact-checking organizations, namely, PolitiFact (politifact.com) and Snopes (snopes.com). We scrape all fact-checked statements, including associated information such as verdicts, fact-checking dates, fact-checkers, and topic tags from these websites (see examples in Table \ref{tab:claim_examples}). As a result, our dataset comprises a total of \num{35014} fact-checked statements, with \num{22561} statements originating from PolitiFact and \num{12453} statements originating from Snopes. The fact-checks have been carried out by \num{608} different fact-checkers. Given that PolitiFact and Snopes have different scales for rating the veracities of the statements, we follow earlier research \cite{vosoughi_spread_2018,markowitz_cross_2023} and classify the ratings into distinct true or false categories.\footnote{Analogous to previous research \cite{markowitz_cross_2023}, we consider fact-checked statements with mixed veracity (\eg, half-true and mixture) as false ratings. The results in our later analysis are qualitatively identical if we exclude statements with mixed veracity (see Suppl. \nameref{sec:no_mixed}).} Specifically, PolitiFact provides six truthfulness ratings, namely, true, mostly true, half true, mostly false, false, and pants on fire. We consider half true, mostly false, false, and pants on fire as false ratings; and the other categories as true ratings. For Snopes, we consider mixture, mostly false, false, miscaptioned, misattributed, scam, and fake as false ratings. True, mostly true, correct attribution, legit, and recall are taken as true ratings.

In addition to the fact-checking verdicts, we clean the fine-grained topic tags provided by the fact-checking organizations and integrate them into our dataset. Specifically, there are in total \num{202} topic tags categorized by the two organizations. The topic tags include many specific social issues, such as, for example, Coronavirus, Guns, and Immigration. Summary statistics of the dataset and most frequent topics are reported in Tables \ref{tab:data_summary} and \ref{tab:topics}.

\subsection{Politicization}

We examine the politicization of fact-checked true and false statements based on the mentions of political elites (see Table \ref{tab:claim_examples}), which is the core of most conceptualizations of politicization \cite{hart_politicization_2020, chinn_politicization_2020}. For this purpose, we gather a comprehensive list of \US political elites based on two sources: (i) we collect \num{4919} public figure names from the People section on the web page of PolitiFact.\footnote{https://www.politifact.com/personalities/} (ii) We additionally collect the names of all \num{1254} members of the \US Congress since 2008 from the official Congress website.\footnote{https://www.congress.gov/members} We merge both data sources together and only consider the political elites affiliated with either the Republican or the Democratic party. The resulting list includes the names of \num{3703} political elites, out of which \num{1938} are Republicans and \num{1765} are Democrats. Note that the list of political elites used in this paper is substantially more comprehensive than the lists used in earlier research \cite{hart_politicization_2020,rathje_out-group_2021,mosleh_measuring_2022}. This allows us to maximize the matches with political elites mentioned in the fact-checked statements.

We use the list of political elites to perform string matching of the fact-checked statements. For this purpose, we first preprocess the texts by removing non-alphabetic and non-numerical characters. Subsequently, we use the names of political elites as keywords to match the cleaned texts. Statements are considered politicized when they mention one or more political elites. Out of \num{35014} fact-checked statements, \num{9308} (26.6\%) are categorized as politicized. Among these, \num{7351} (79\%) statements mention only one politician, and \num{1957} (21\%) statements mention two or more politicians. Notably, only \num{674} (7.2\%) statements mention Democrats and Republicans at the same time. This suggests that most of the politicized statements have a specific focus on either Democrats or Republicans. 

\begin{figure}
    \centering
    \includegraphics[width=.45\textwidth]{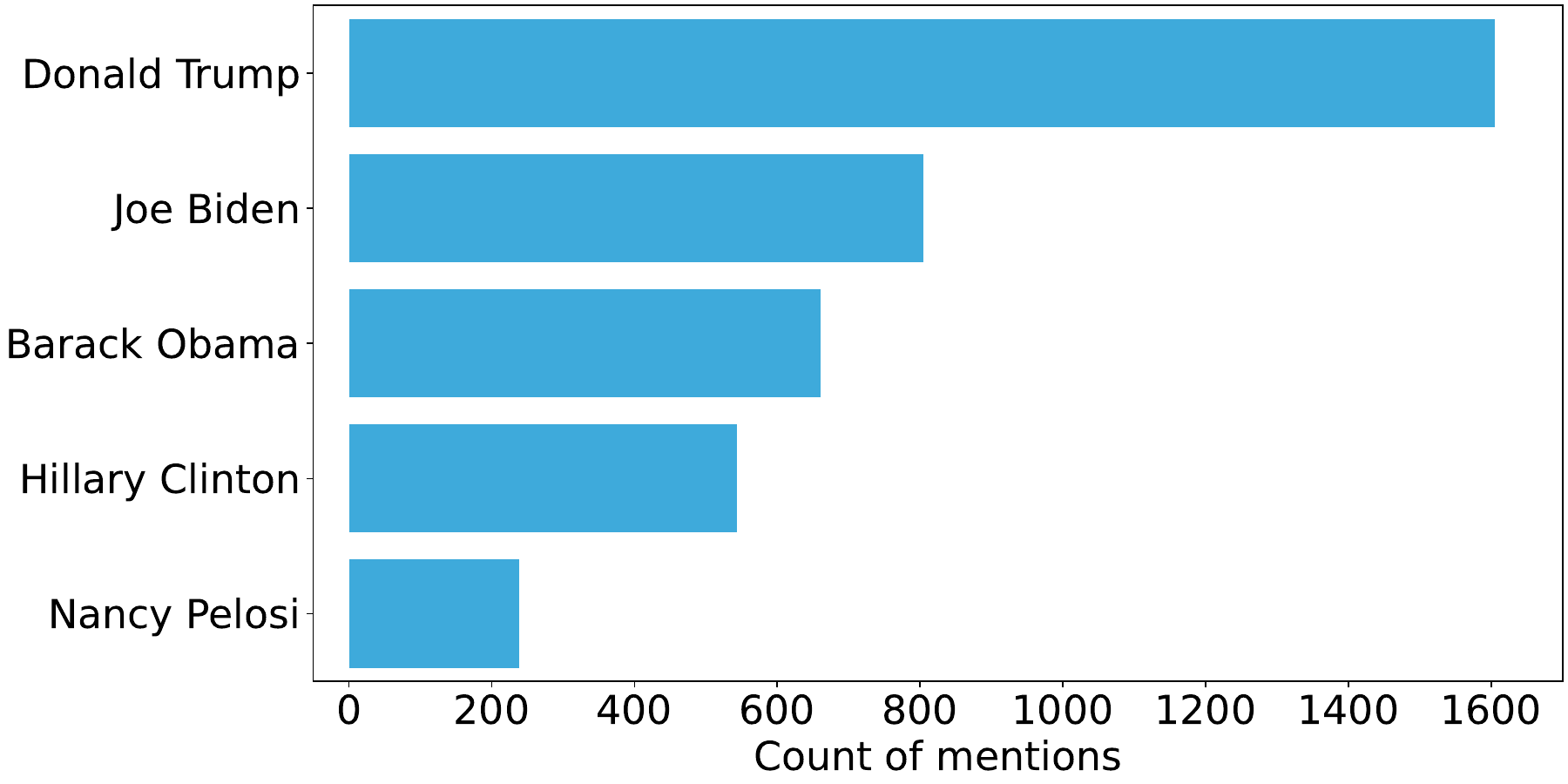}
    \caption{The five most frequently mentioned politicians in fact-checked statements.}
    \label{fig:count_mentions}
\end{figure}

Fig. \ref{fig:count_mentions} shows that Donald Trump, Joe Biden, Barack Obama, Hillary Clinton, and Nancy Pelosi are the five most frequently mentioned politicians in the fact-checked statements. The number of fact-checked statements that mention one of these five politicians is \num{3670}, which accounts for 39.4\% of all politicized statements.

\textbf{Manual validation:} To evaluate the performance of the string matching approach, we manually assess \num{200} fact-checked statements randomly selected from our dataset. The weighted average scores of precision and recall are 91\% and 89\%, respectively, which indicates a high performance of the string-matching approach. As a further check, we evaluate all politicized statements by additionally extracting fact-checking comments, \ie, the textual verdicts of fact-checkers, for the corresponding statements (see Suppl. \nameref{sec:manual_val} for details). Here, we find an almost perfect precision score of 99.6\%. 

\textbf{Partisan scores:}
Additionally, each political elite has a distinct political affiliation, namely, Republican or Democratic. We code the statements according to the political affiliations of the political elites mentioned in the textual content. If the political elite is Democrat, the partisan score is \num{-1}, and if the political elite is Republican, the partisan score is 1. If the statements contain the names of more than one political elite, we sum the partisan scores. For example, if a statement mentions one Republican and one Democrat, then the statement has a partisan score of 0. A positive score ($>0$) indicates that the statement predominantly mentions Republicans, whereas a negative score ($<0$) signifies a predominant mentioning of Democrats. Out of all fact-checked statements, \num{4403} statements predominantly mention Republicans, and \num{4455} statements predominantly mention Democrats.

\subsection{Other-Condemning Emotions} 

Other-condemning emotions are a subset of moral emotions and comprise the emotions contempt, anger, and disgust \cite{solovev_moral_2022,dastani_other-condemning_2017}. We use a dictionary-based approach to computationally measure the expression of other-condemning emotions. Specifically, we employ the NRC emotion lexicon to match emotion words in the cleaned text of the fact-checked statements. The NRC emotion lexicon is a widely used tool in previous research \cite{vosoughi_spread_2018,prollochs_emotions_2021,solovev_moral_2022,chuai_anger_2022}. It includes \num{14154} unique English words and their associations with eight basic emotions \cite{mohammad_crowdsourcing_2013}. Additionally, the eight basic emotions, namely, anger, anticipation, disgust, fear, joy, sadness, surprise, and trust, can be combined into 24 complex emotional dyads \cite{plutchik_emotions_1991}. For example, contempt is composed of anger and disgust. In line with the prior work that has validated the effectiveness of using NRC emotion lexicon to measure other-condemning emotions \cite{solovev_moral_2022}, we calculate the scores of the eight basic emotions and the 24 derived emotions based on the frequency of words belonging to each emotion category. Subsequently, the score of other-condemning emotions is calculated as the sum of the scores for contempt, anger, and disgust. The mean score for other-condemning emotions in the fact-checked statements is \num{0.096} (see Table \ref{tab:data_summary} for details). This indicates that, on average, the three other-condemning emotions account for 9.6\% of all emotion components from Plutchik's emotion model.

\subsection{Explanatory Regression Models}
In this study, we employ explanatory regression modeling to answer our research questions. The variables included in our models are as follows (summary statistics are in Table \ref{tab:data_summary}):
\begin{itemize}[leftmargin=*]
    \item $\bm{\var{Politician}}$: A dummy variable indicating that the fact-checked statement explicitly mentions politician(s) ($=$1; otherwise 0).
    \item $\bm{\var{Republican}}$: A dummy variable indicating that the fact-checked statement predominantly mentions Republicans ($=$1; otherwise 0). 
    \item $\bm{\var{Democrat}}$: A dummy variable indicating that the fact-checked statement predominantly mentions Democrats ($=$1; otherwise 0).
    \item $\bm{\var{Falsehood}}$: A dummy variable indicating that the fact-checked statement is labeled as false ($=$1; otherwise 0).
    \item $\bm{\var{MTE}}$ (Months to Elections): A continuous variable measuring the time interval (in months) between the date of each fact-checked statement and the next upcoming election. Here, we consider 2-year election cycles including both \US presidential and midterm elections.
    \item $\bm{\var{Words}}$: The number of words in the cleaned text of each fact-checked statement. This (control) variable is commonly used to account for the content richness of the statements \cite{miani_interconnectedness_2022}.
    \item $\bm{\var{OtherCondemning}}$: A continuous variable measuring other-condemning emotions in fact-checked statements.
\end{itemize}

To address research questions RQ1, RQ2, and RQ3, we treat $\var{Politician}$, $\var{Republican}$, and $\var{Democrat}$ as dependent variables. The explanatory variables are $\var{Falsehood}$, $\var{MTE}$, and $\var{Words}$. Note that some fact-checkers might be more focused on politics and, thus, may prefer to check statements that mention political elites. Hence, we control for such heterogeneity among fact-checkers through the use of fact-checker-specific random effects (\ie, random intercepts). Formally, we specify the following three logistic regression models (one for each of the three dependent variables):
\begin{equation}
\label{equ:logit}
\begin{aligned}
    logit(y) = & \, \beta_{0} + \beta_{1}Falsehood_{i} + \beta_{2}MTE_{i} \\
    & + \beta_{3}Falsehood_{i} \times MTE_{i} + \beta_{4}Words_{i} \\ 
   & + u_{\text{year}} + u_{\text{factchecker}},
\end{aligned}
\end{equation}
where $y$ is the binary dependent variable (i.e., $\var{Politician}$, $\var{Republican}$, or $\var{Democrat}$), $\beta_{0}$ is the intercept, $\beta_{1}$ to $\beta_{4}$ are the coefficients of the explanatory variables, $u_{\text{year}}$ are yearly fixed effects, and $u_{\text{factchecker}}$ are random effects specific to the fact-checkers. Notably, we also include an interaction term $\var{Falsehood_{i} \times MTE_{i}}$, which allows us to analyze whether the level of politicization in fact-checked false statements differs from that in fact-checked true statements during election periods (RQ3).

To address RQ4, we consider $\var{OtherCondemning}$ as the dependent variable, while treating $\var{Politician}$ as an independent variable. The rationale is that the politicization of statements may be linked to the expression of other-condemning emotions \cite{rathje_out-group_2021,dastani_other-condemning_2017}. Consequently, we specify the following linear regression model with fact-checker-specific random effects and yearly fixed effects:
\begin{equation}
\label{equ:linear}
\begin{aligned}
    OtherCondemning = & \, \beta_{0} + \beta_{1}Politician_{i} \\
    &+ \beta_{2}Falsehood_{i}\\
    &+ \beta_{3}MTE_{i} + \beta_{4}Words_{i} \\
    &+ u_{\text{year}} + u_{\text{factchecker}}.
\end{aligned}
\end{equation}
Note that in our later analysis, we also consider the interactions of $\var{Politician}$ with both $\var{Falsehood}$ and $\var{MTE}$. This allows us to investigate potential variations concerning the politicization of true and false fact-checked statements and how the effects vary during elections.

\section{Empirical Analysis}

\begin{figure*}[htb]
     \centering
     \begin{subfigure}[b]{0.32\textwidth}
         \centering
         \includegraphics[width=\textwidth]{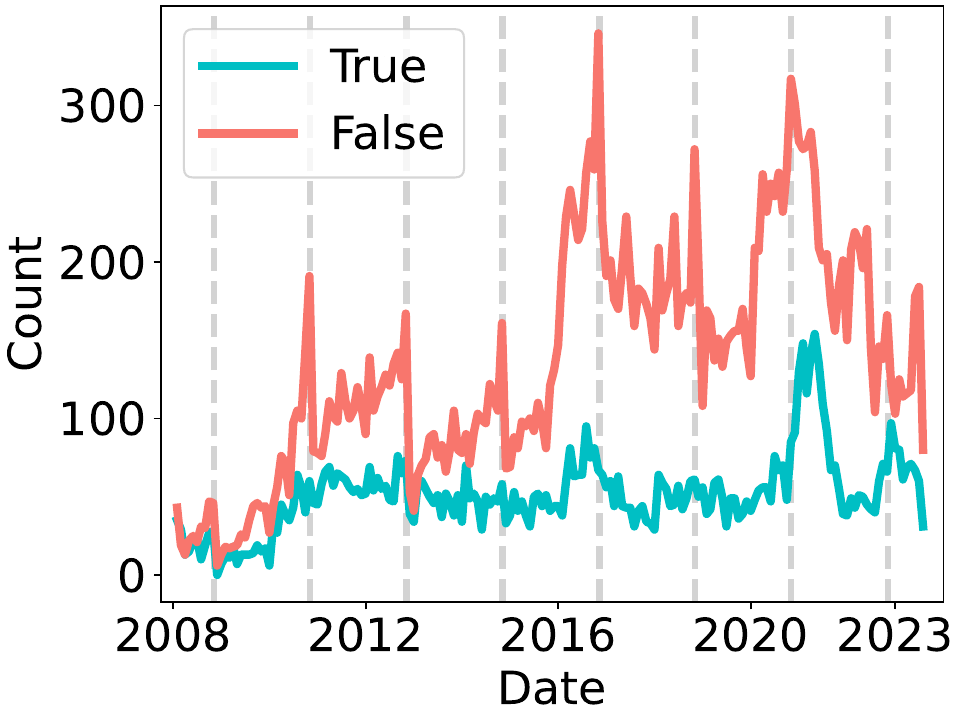}
         \caption{}
         \label{fig:claim_count_by_time}
     \end{subfigure}
     \hfill
     \begin{subfigure}[b]{0.32\textwidth}
         \centering
         \includegraphics[width=\textwidth]{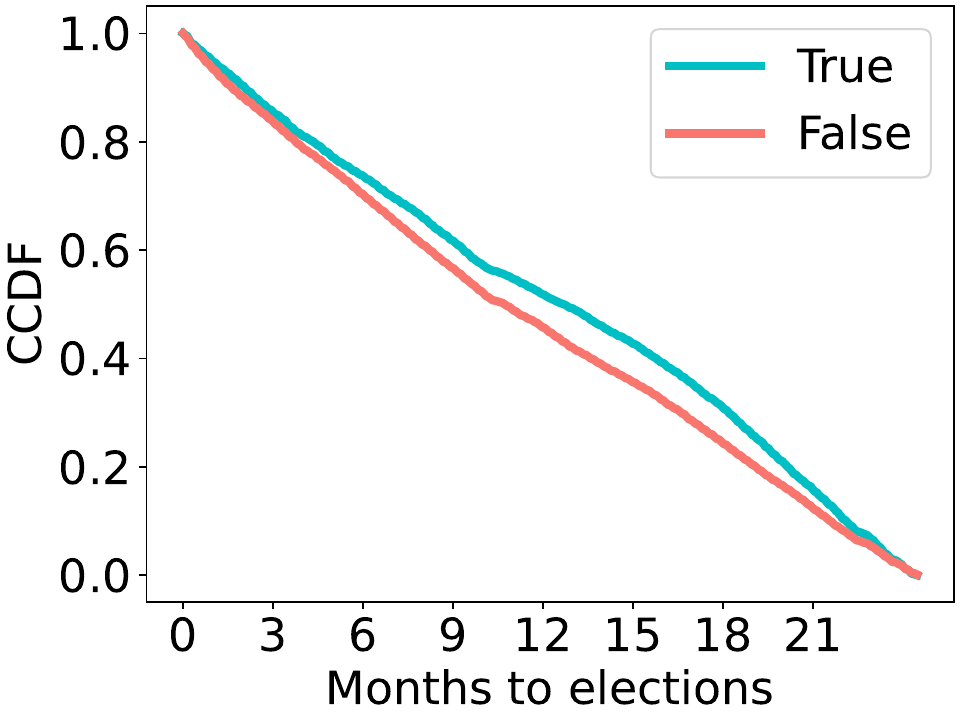}
         \caption{}
         \label{fig:months_elections_ccdf}
     \end{subfigure}
     \hfill
     \begin{subfigure}[b]{0.32\textwidth}
         \centering
         \includegraphics[width=\textwidth]{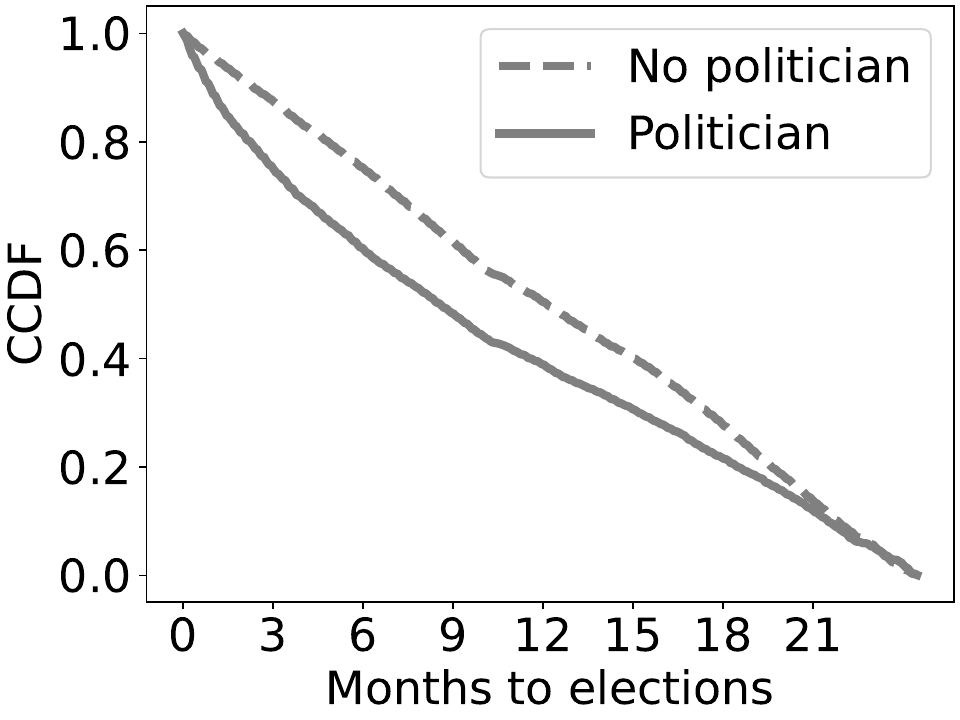}
         \caption{}
         \label{fig:months_elections_politician_ccdf}
     \end{subfigure}
     \caption{Dataset overview. \subref{fig:claim_count_by_time} The monthly count of fact-checked true and false statements in our observation period between 2008 and 2023. The eight grey vertical dash lines correspond to eight \US election dates. \subref{fig:months_elections_ccdf} Complementary Cumulative Distribution Functions (CCDFs) for $\var{MTE}$ (months to elections) in fact-checked true and false statements. The difference in distributions is statistically significant according to a KS-test ($\var{KS} = 0.073, p < 0.001$). \subref{fig:months_elections_politician_ccdf} CCDFs for $\var{MTE}$ in fact-checked statements w/ and w/o mentions of political elites. The difference in distributions is statistically significant according to a KS-test ($\var{KS} =0.151, p < 0.001$).}
    \label{fig:summary_statistics}
\end{figure*}

Our dataset consists of \num{35014} true and false statements that have been fact-checked by major \US fact-checking organizations between January 2008 and July 2023. The dataset thus spans an observation period of approximately 15 years, including eight \US Midterm and Presidential elections. The number of fact-checked false statements tends to increase when approaching elections, with the highest peak observed during the 2016 \US Presidential election (Fig. \ref{fig:claim_count_by_time}). In contrast, the number of fact-checked true statements is relatively stable with a significant peak after the 2020 \US Presidential election. Fig. \ref{fig:months_elections_ccdf} shows that the mean of the variable Months to Elections ($\var{MTE}$) in fact-checked false statements (mean of \num{11.321}) is smaller than that in fact-checked true statements (mean of \num{12.349}). This implies that fact-checks for false statements are more concentrated close to elections than those for true statements. Furthermore, the mean of $\var{MTE}$ in fact-checked statements mentioning political elites (mean of \num{10.031}) is smaller than that in fact-checked statements without mentioning political elites (mean of \num{12.173}). This suggests that as elections approach, fact-checkers are more likely to target politicized statements over non-politicized ones (Fig. \ref{fig:months_elections_politician_ccdf}).

\subsection{Mentions of Political Elites (RQ1 \& RQ2)}

We estimate logistic regression models to analyze how veracity ratings are linked to mentions of \US political elites (Republicans and Democrats) in the fact-checked statements (RQ1 \& RQ2). The regression results are reported in Table \ref{tab:politicians}. Columns (1), (2), and (3) correspond to the dependent variables $\var{Politician}$, $\var{Republican}$, and $\var{Democrat}$, respectively. As shown in Column (1), the coefficient estimate for $\var{Falsehood}$ is statistically significant and positive ($\var{coef.}=0.182, p<0.001$), with an odds ratio ($\var{o.r.}$) of $e^{0.182} = 1.200$ (95\% CI: $[1.083, 1.330]$). This implies that false statements are 20.0\% more likely to mention political elites, as compared to true statements (RQ1). 

Subsequently, we examine how the mentions of Republicans and Democrats differ between fact-checked true and false statements. For $\var{Republican}$ [Column (2)], we find that the coefficient estimate for $\var{Falsehood}$ is significantly negative ($\var{coef.}=-0.308$, $p<0.001$; $\var{o.r.}=0.735$, 95\% CI: $[0.648, 0.833]$). This implies that false statements are 26.5\% less likely to mention Republicans compared to true statements. Conversely, when examining $\var{Democrat}$ [Column (3)], we find a statistically significant positive coefficient estimate for $\var{Falsehood}$ ($\var{coef.}=0.632$, $p<0.001$; $\var{o.r.}=1.881$, 95\% CI: $[1.630, 2.172]$). This indicates that fact-checked false statements are 88.1\% more likely to mention Democrats than fact-checked true statements. Additionally, the higher likelihood of mentioning Democrats in fact-checked false statements compared to fact-checked true statements is robust across topics (see Suppl. \nameref{sec:analysis_across_topics} for details). In summary, our findings imply that fact-checked false statements are more likely to mention Democrats and less likely to mention Republicans than fact-checked true statements (RQ2).

\begin{table}[htb]
\footnotesize
\begin{center}
\setlength{\tabcolsep}{4pt}
\begin{tabularx}{\columnwidth}{@{\hspace{\tabcolsep}\extracolsep{\fill}}l*{3}{S}}
\toprule
&\multicolumn{1}{c}{(1)}&\multicolumn{1}{c}{(2)}&\multicolumn{1}{c}{(3)}\\
&\multicolumn{1}{c}{Politician}&\multicolumn{1}{c}{Republican}&\multicolumn{1}{c}{Democrat}\\
\midrule
Falsehood   &       0.182\sym{***}&      -0.308\sym{***}&       0.632\sym{***}\\
            &     (0.052)         &     (0.064)         &     (0.073)         \\
MTE         &      -0.037\sym{***}&      -0.033\sym{***}&      -0.033\sym{***}\\
            &     (0.004)         &     (0.005)         &     (0.006)         \\
Falsehood $\times$ MTE&       0.007         &       0.018\sym{***}&      -0.004         \\
            &     (0.004)         &     (0.005)         &     (0.006)         \\
Words       &       0.022\sym{***}&       0.019\sym{***}&       0.014\sym{***}\\
            &     (0.002)         &     (0.002)         &     (0.002)         \\
Intercept      &      -1.225\sym{***}&      -2.157\sym{***}&      -2.119\sym{***}\\
            &     (0.134)         &     (0.172)         &     (0.153)         \\
\midrule
Fact-checker REs       &       \checkmark&       \checkmark&       \checkmark\\
Year FEs     &       \checkmark&       \checkmark&       \checkmark\\
\midrule
\#Statements       &       \num{35014}         &       \num{35014}         &       \num{35014}         \\
\bottomrule
\end{tabularx}
\caption{Estimation results for mixed logistic regression models predicting whether the fact-checked statements mention political elites of either political party [Column (1)], Republicans [Column (2)], or Democrats [Column (3)]. Fact-checker-specific random effects (REs) and yearly fixed effects (FEs) are included. Reported are coefficient estimates with standard errors in parentheses. \sym{*} \(p<0.05\), \sym{**} \(p<0.01\), \sym{***} \(p<0.001\).
}
\label{tab:politicians}
\end{center}
\end{table}

\textbf{Analysis across incumbent \US presidents:}
The observation period covered by our dataset includes three full presidential cycles, \ie, two presidential terms from Barack Obama (from 2008-11-04 through 2012-11-06 to 2016-11-08, Obama's incumbency) and one presidential term from Donald Trump (from 2016-11-08 to 2020-11-03, Trump's incumbency).\footnote{Additionally, our dataset covers two years of the presidency of Joe Biden. To ensure comparability with other presidents' incumbencies, we omit the period of Biden's incumbency and analyze the temporal trend of mentions of political elites during the period from 2008 to 2020 and across Obama's and Trump's incumbencies.} Specifically, there are \num{14515} fact-checked statements during Obama's incumbency and \num{11452} fact-checked statements during Trump's incumbency, respectively. We incorporate an incumbency fixed effect ($\var{Trump}$) and its interaction with $\var{Falsehood}$ into the regression models to examine whether there is a difference in mentioning political elites in fact-checked statements between Obama's incumbency and Trump's incumbency. 

According to the estimation results [see Column (1) of Table \ref{tab:politician_incumbency_fixed}], the coefficient estimate of $\var{Trump}$ is significantly positive ($\var{coef.}=0.174$, $p<0.05$; $\var{o.r.}=1.190$, 95\% CI: $[1.039, 1.363]$). This indicates that fact-checked statements (both true and false) during Trump's incumbency, were 19\% more likely to mention political elites (either Democrats or Republicans) compared to those during Obama's incumbency. However, when we only consider mentions of Republicans [see Column (2) of Table \ref{tab:politician_incumbency_fixed}], the coefficient estimate of $\var{Falsehood \times Trump}$ is significantly negative ($\var{coef.}=-0.258$, $p<0.01$; $\var{o.r.}=0.772$, 95\% CI: $[0.651, 0.916]$). This implies that fact-checked false statements during Trump's incumbency were 22.8\% less likely to mention Republicans than those during Obama's incumbency. Overall, this indicates a more pronounced partisan asymmetry in fact-checking activity during Trump's incumbency compared to Obama's incumbency.

\begin{figure}[htb]
    \centering
    \includegraphics[width=.45\textwidth]{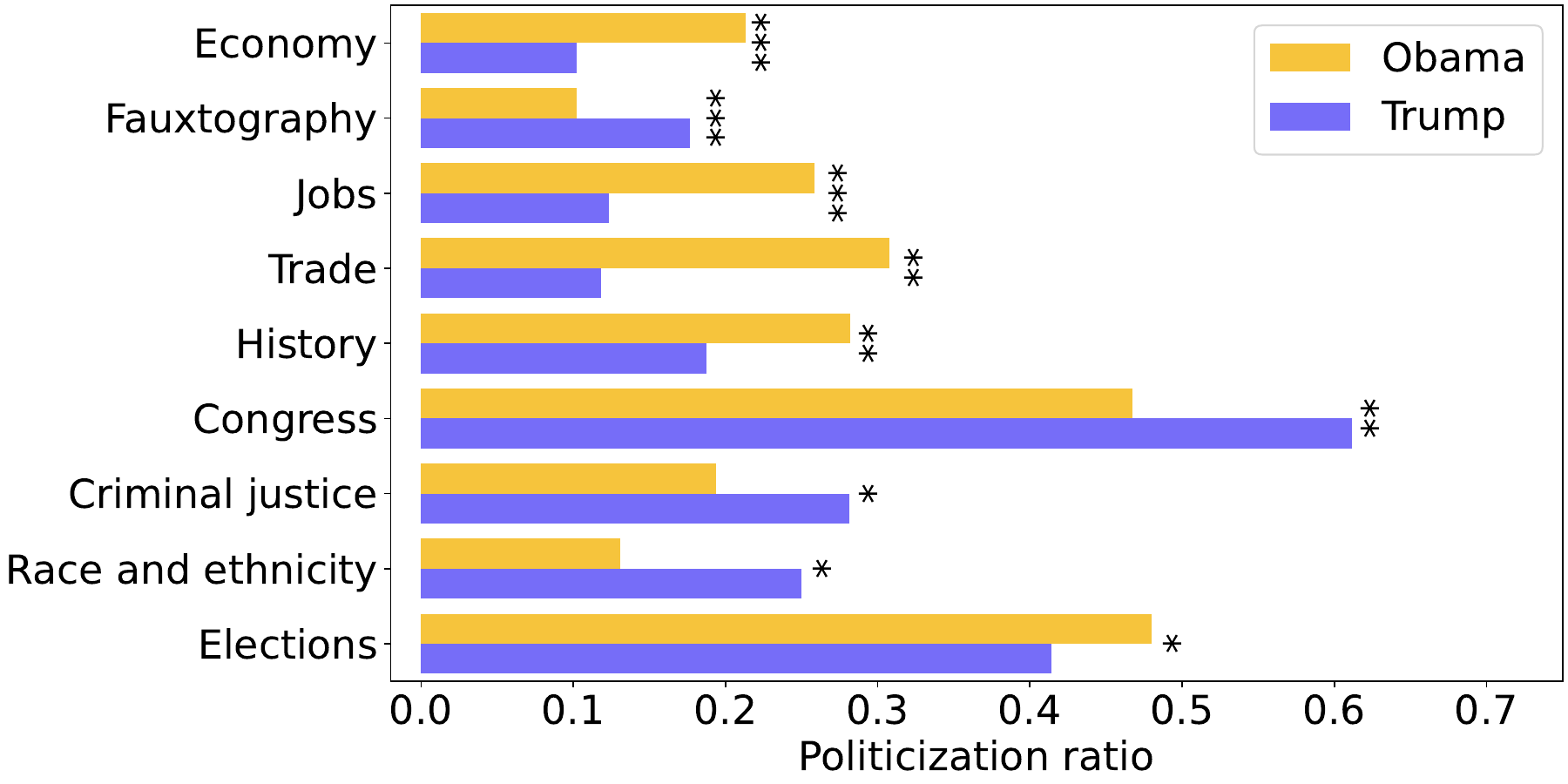}
    \caption{The politicization ratios of the nine topics that have statistically significant differences during Obama's and Trump's incumbencies. The Chi-Square test is conducted for each topic group. \sym{*} \(p<0.05\), \sym{**} \(p<0.01\), \sym{***} \(p<0.001\).}
    \label{fig:politi_issues_by_incumbency}
\end{figure}

We further examine how the politicization of individual topics (\eg, Economy, Race and Ethnicity, etc.) has changed across presidential incumbencies. To this end, we calculate a politicization ratio for each of the topics measuring the proportion of statements mentioning political elites relative to all fact-checked statements on the same topic. We find that the majority of topics have similar politicization ratios between Obama's incumbency and Trump's incumbency, with only nine topics having significant differences. Specifically, as shown in Fig. \ref{fig:politi_issues_by_incumbency}, the topics of ``Economy,'' ``Jobs,'' ``Trade,'' ``History,'' and ``Elections'' have higher politicization ratios during Obama's incumbency, while the topics of ``Fauxtography,'' ``Congress,'' ``Criminal justice,'' and ``Race and ethnicity'' have higher politicization ratios during Trump's incumbency (see Suppl. \nameref{sec:analysis_across_topics} for details).

\subsection{Elections (RQ3)}
To answer RQ3, we examine how the likelihood of mentions of political elites in fact-checked statements varies with regard to the time distance to \US elections. The coefficient estimates for the variable $\var{MTE}$ (see Table \ref{tab:politicians}) are consistently negative and statistically significant [$\var{coef.}=-0.037$, $p<0.001$, $\var{o.r.}=0.963$, 95\% CI: $[0.956, 0.970]$ in Column (1); $\var{coef.}=-0.033$, $p<0.001$, $\var{o.r.}=0.967$, 95\% CI: $[0.958, 0.976]$ in Column (2); and $\var{coef.}=-0.033$, $p<0.001$, $\var{o.r.}=0.968$, 95\% CI: $[0.957, 0.978]$ in Column (3)]. This suggests that fact-checked statements are more likely to mention political elites (both Republicans and Democrats) as elections approach. Specifically, one month closer to elections corresponds to a 3.7\% increase in the odds of mentioning political elites of either Republicans or Democrats in fact-checked statements. Furthermore, the coefficient estimate for the interaction $\var{Falsehood \times MTE}$ in Column (2) of Table \ref{tab:politicians} is significantly positive ($\var{coef.}=0.018$, $p<0.001$; $\var{o.r.}=1.018$, 95\% CI: $[1.008, 1.028]$). This means that, as time progresses away from elections, the likelihood of mentioning Republicans in fact-checked false statements decreases at a slower rate compared to that in fact-checked true statements. However, the coefficient estimate for $\var{Falsehood \times MTE}$ in Column (3) of Table \ref{tab:politicians} is not statistically significant ($\var{coef.}=-0.004$, $p>0.05$; $\var{o.r.}=0.996$, 95\% CI: $[0.985, 1.007]$), which indicates that there is no corresponding significant difference in the changes in mentions of Democrats before elections between fact-checked true and false statements. 

\begin{figure*}[htb]
     \centering
     \begin{subfigure}[b]{0.32\textwidth}
         \centering
         \includegraphics[width=\textwidth]{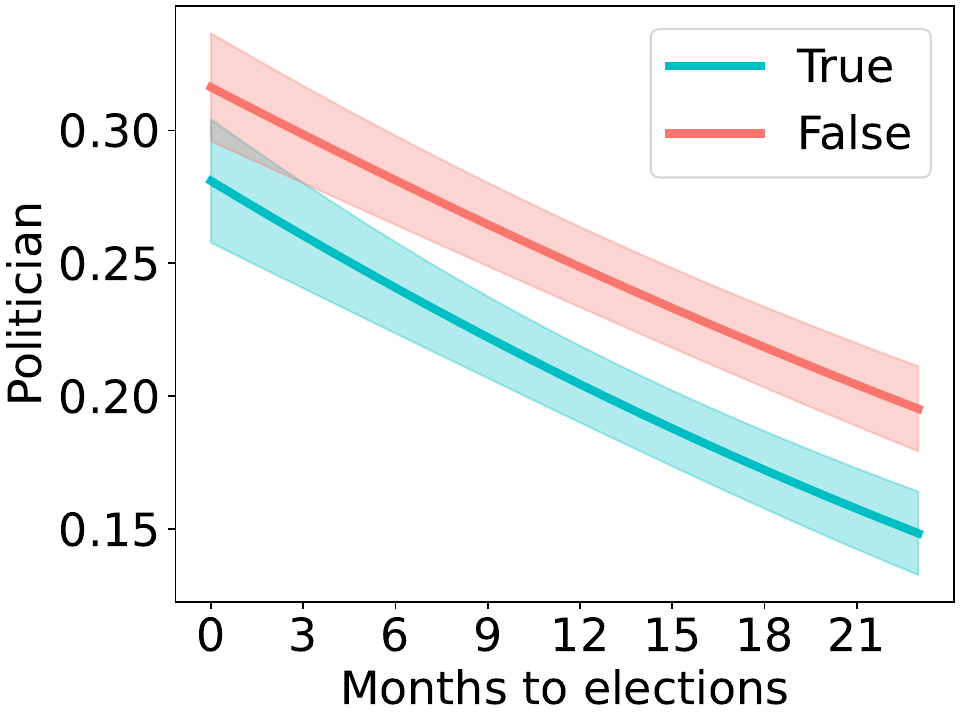}
         \caption{}
         \label{fig:politician_margin}
     \end{subfigure}
     \hfill
     \begin{subfigure}[b]{0.32\textwidth}
         \centering
         \includegraphics[width=\textwidth]{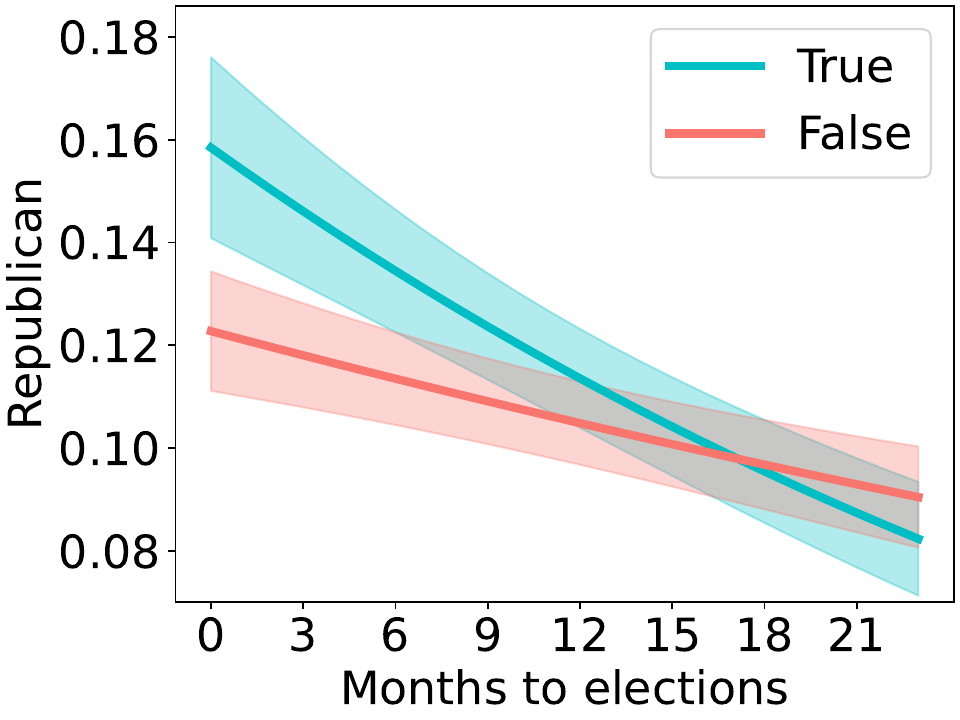}
         \caption{}
         \label{fig:republican_margin}
     \end{subfigure}
     \hfill
     \begin{subfigure}[b]{0.32\textwidth}
         \centering
         \includegraphics[width=\textwidth]{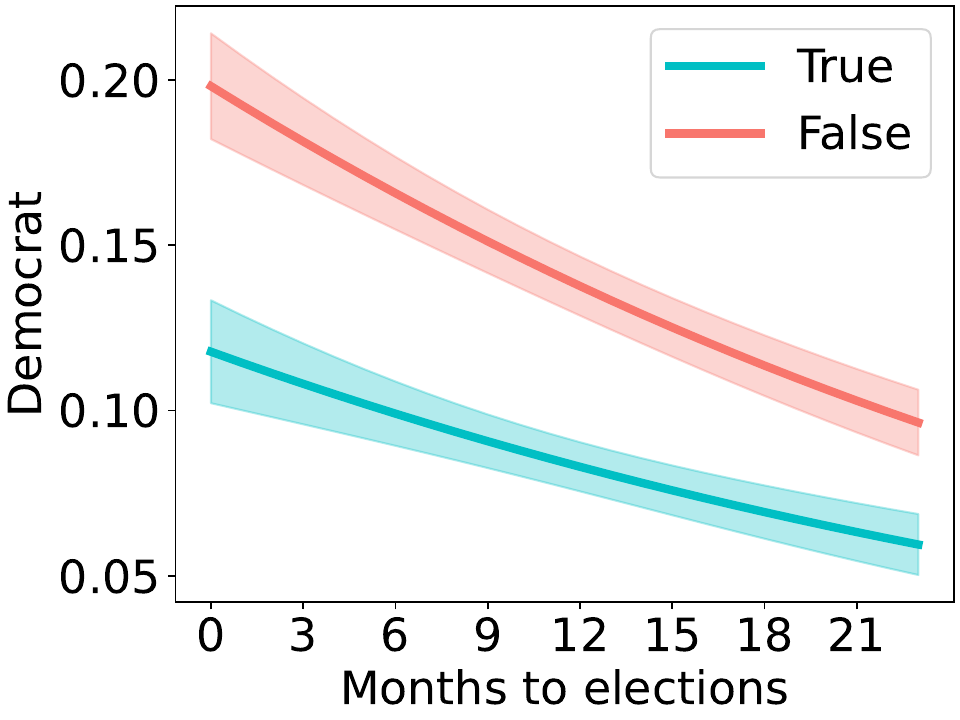}
         \caption{}
         \label{fig:democrat_margin}
     \end{subfigure}
    \caption{Predicted marginal means for the probabilities of mentions of \subref{fig:politician_margin} political elites of either political party, \subref{fig:republican_margin} Republicans, and \subref{fig:democrat_margin} Democrats in fact-checked true and false statements. The error bands represent 95\% Confidence Intervals (CIs).}
    \label{fig:politician_margins}
\end{figure*}

The predicted marginal means\footnote{Within our logistic regression models, the predicted marginal means indicate the probability of mentioning political elites in fact-checked statements.} for fact-checked true and false statements following elections are visualized in Fig. \ref{fig:politician_margins}. Consistent with our assessment of the coefficient estimates, fact-checked statements are more likely to mention politicians and especially Democrats, with the probability of mentioning political elites increasing in parallel for both true and false statements when elections approach (RQ3).

\subsection{Other-Condemning Emotions (RQ4)}

We now analyze whether fact-checked statements that mention political elites express more other-condemning emotions compared to fact-checked statements that do not mention political elites (RQ4). The regression results are reported in Table \ref{tab:emotions}. We first examine a baseline model without interactions [Column (1)]. The coefficient estimate for $\var{Politician}$ is significantly positive ($\var{coef.}=0.009$, $p<0.001$). Specifically, the predicted marginal means for other-condemning emotions in politicized and non-politicized statements are \num{0.102} (95\% CI: $[\num{0.097}, \num{0.106}]$) and \num{0.093} (95\% CI: $[\num{0.090}, \num{0.096}]$), respectively. This implies that fact-checked statements with political elites express 9.7\% more other-condemning emotions compared to those without political elites (RQ4). Additionally, the significantly positive coefficient estimate for $\var{Falsehood}$ suggests that fact-checked false statements contain more other-condemning emotions compared to fact-checked true statements ($\var{coef.}=0.007, p<0.01$). The predicted marginal means for other-condemning emotions in fact-checked true and false statements are \num{0.090} (95\% CI: $[\num{0.086}, \num{0.094}]$) and \num{0.097} (95\% CI: $[\num{0.094}, \num{0.100}]$), respectively. This implies that fact-checked false statements express 7.8\% more other-condemning emotions compared to fact-checked true statements. Notably, the coefficient estimate for $\var{MTE}$ is not statistically significant ($\var{coef.}=0.000, p>0.05$), indicating that the expression of other-condemning emotions is not significantly linked to time to elections. 

We further examine the moderating effects of the variables $\var{Falsehood}$ and $\var{MTE}$ on $\var{Politician}$. As shown in Column (2) of Table \ref{tab:emotions}, the coefficient estimate for the interaction $\var{Politician \times Falsehood}$ is not statistically significant ($\var{coef.}=-0.004, p>0.05$), while the coefficient estimate for $\var{Falsehood}$ is significantly positive ($\var{coef.}=0.008, p<0.01$). This suggests that fact-checked false statements carry more other-condemning emotions compared to fact-checked true statements, regardless of whether political elites are mentioned. In contrast, the coefficient estimate for $\var{Politician \times MTE}$ is significantly negative ($\var{coef.}=-0.001, p<0.01$), whereas the coefficient estimate for $\var{MTE}$ is not significant ($\var{coef.}=0.000, p>0.05$). This suggests that the expression of other-condemning emotions in fact-checked statements mentioning political elites decreases as the time period moves further away from elections. 

Additionally, we include $\var{Republican}$ and $\var{Democrat}$ in the model [Column (3) of Table \ref{tab:emotions}]. However, we find that the coefficient estimates for the two variables are both not statistically significant (each $p>0.05$). This indicates that mentioning political elites, either Republican or Democrat, has a similarly positive association with other-condemning emotions within fact-checked statements.

\textbf{Analysis across incumbent \US presidents:}
To analyze differences in other-condemning emotions of fact-checked statements between Obama's and Trump's incumbencies, we incorporate incumbency fixed effect ($\var{Trump}$) and its interactions with $\var{Politician}$ and $\var{Falsehood}$ into the base model. According to the estimation results [see Column (1) of Table \ref{tab:emotions_incumbency}], the coefficient estimate of $\var{Trump}$ is significantly positive ($\var{coef.}=0.014, p<0.01$), which indicates that other-condemning emotions in fact-checked statements during Trump's incumbency are higher than that during Obama's incumbency. Moreover, the coefficient estimates of $\var{Politician \times Trump}$ ($\var{coef.}=-0.008, p>0.05$) and $\var{Falsehood \times Trump}$ ($\var{coef.}=0.005, p>0.05$) are not statistically significant. This suggests that the high other-condemning emotions during Trump's incumbency are not specific to politicized or false statements but general to all fact-checked statements.

\begin{table}[htb]
\footnotesize
\begin{center}
\setlength{\tabcolsep}{0pt}
\begin{tabularx}{\columnwidth}{@{\hspace{\tabcolsep}\extracolsep{\fill}}l*{3}{S}}
\toprule
&\multicolumn{1}{c}{(1)}&\multicolumn{1}{c}{(2)}&\multicolumn{1}{c}{(3)}\\
&\multicolumn{1}{c}{Baseline}&\multicolumn{1}{c}{Interactions}&\multicolumn{1}{c}{Partisanship}\\
\midrule
Politician  &       0.009\sym{***}&       0.020\sym{***}&       0.025\sym{*}  \\
            &     (0.002)         &     (0.006)         &     (0.010)         \\
Republican  &                     &                     &      -0.005         \\
            &                     &                     &     (0.009)         \\
Democrat    &                     &                     &      -0.004         \\
            &                     &                     &     (0.009)         \\
Falsehood     &       0.007\sym{**} &       0.008\sym{**} &       0.008\sym{**} \\
            &     (0.002)         &     (0.003)         &     (0.003)         \\
Politician $\times$ Falsehood &                     &      -0.004         &      -0.004         \\
            &                     &     (0.005)         &     (0.005)         \\
MTE         &      -0.000         &       0.000         &       0.000         \\
            &     (0.000)         &     (0.000)         &     (0.000)         \\
Politician $\times$ MTE   &                     &      -0.001\sym{**} &      -0.001\sym{**} \\
            &                     &     (0.000)         &     (0.000)         \\
Words       &       0.001\sym{***}&       0.001\sym{***}&       0.001\sym{***}\\
            &     (0.000)         &     (0.000)         &     (0.000)         \\
Intercept      &       0.099\sym{***}&       0.095\sym{***}&       0.095\sym{***}\\
            &     (0.009)         &     (0.009)         &     (0.009)         \\
\midrule
Fact-checker REs       &       \checkmark&       \checkmark&       \checkmark\\
Year FEs       &    \checkmark       &     \checkmark      &     \checkmark       \\
\midrule
\#Statements       &       \num{35014}         &       \num{35014}         &       \num{35014}         \\
\bottomrule
\end{tabularx}
\caption{Estimation results for mixed linear regression models predicting other-condemning emotions in the fact-checked statements. Columns (1)--(3) report the estimation results from the baseline model, model with interactions, and model with party indicators of ($\var{Republican}$ and $\var{Democrat}$), respectively. Fact-checker-specific random effects and yearly fixed effects are included. Reported are coefficient estimates with standard errors in parentheses. \sym{*} \(p<0.05\), \sym{**} \(p<0.01\), \sym{***} \(p<0.001\).}
\label{tab:emotions}
\end{center}
\end{table}

\subsection{Exploratory Analysis \& Robustness Checks}
\label{sec:exploratory_robustness}

\textbf{Partisan leanings of fact-checked statements:}
Our study suggests that fact-checked false statements are more likely to mention Democrats compared to fact-checked true statements. However, previous research indicates that widely shared fact-checked false statements tend to be pro-Republican (\ie, benefiting the political right, right-leaning) \cite{garrett_conservatives_2021,chen_neutral_2021}. This dissonance may possibly be reconciled by fact-checked false statements being more likely to harness animosity toward Democrats compared to true statements. To test this notion, we further examine the use of other-condemning emotions toward political elites and explore the partisan leanings of fact-checked true and false statements. Here, we combine mentions of politicians (Republicans or Democrats) and other-condemning emotions together to define a new dependent variable $\var{PartisanLeaning}$. We code fact-checked statements as right-leaning (pro-Republican) if they utilize other-condemning emotions within fact-checked statements that predominantly mention Democrats; and vice versa. Formally, we define the dependent variable $\var{PartisanLeaning}$ as $- \var{PartisanScore} \times \var{OtherCondemning}$, where $\var{PartisanLeaning}>0$ means that the statement is right-leaning (pro-Republican), and $\var{PartisanLeaning}<0$ means that the statement is left-leaning (pro-Democrat). 

Subsequently, we estimate mixed linear regressions predicting partisan leanings in fact-checked true and false statements (see Suppl. \nameref{sec:political_leanings} for details and full estimation results). For all fact-checked statements, the coefficient estimate of the explanatory variable $\var{Falsehood}$ is significantly positive [$\var{coef.}=0.009, p<0.001$; see Column (1) of Table \ref{tab:partisan_leaning}]. This indicates that fact-checked false statements are more pro-Republican and use higher other-condemning emotions when mentioning Democrats, compared to true statements. The predicted marginal means for partisan leanings in true and false statements are \num{-0.005} (95\% CI: $[\num{-0.007}, \num{-0.002}]$) and \num{0.004} (95\% CI: $[\num{0.002}, \num{0.006}]$), respectively. This indicates that, on average, fact-checked false statements are pro-Republican, whereas fact-checked true statements are pro-Democrat.

Furthermore, PolitiFact provides author information for the fact-checked statements. This allows us to partially identify the political elites who authored the fact-checked statements based on our politician database (see Suppl. \nameref{sec:political_leanings} for details). Specifically, there are \num{13693} fact-checked statements from political elites, among which \num{5640} from Democrats and \num{8053} from Republicans. Given this, we conduct a subgroup analysis for fact-checked statements from PolitiFact and validate whether politicians express more other-condemning emotions in the fact-checked statements that mention political out-groups compared to non-partisan authors. According to the estimation results [see Column (3) of Table \ref{tab:partisan_leaning}], fact-checked statements from Democrats are more pro-Democrat than those from non-Democrats ($\var{coef.}=-0.021, p<0.001$). In contrast, fact-checked statements from Republicans are more pro-Republican than those from non-Republicans ($\var{coef.}=0.012, p<0.001$). Further details and full estimation results are provided in Suppl. \nameref{sec:political_leanings}.

Taken together, concordant with our hypothesis, we find that fact-checked false statements are indeed more pro-Republican than true fact-checked statements. Furthermore, fact-checked false statements from Republicans are more pro-Republican and express higher other-condemning emotions when mentioning Democrats compared to fact-checked true statements; and we observe the same pattern for fact-checked false statements from Democrats. These results strengthen the interpretation of our findings and suggest that fact-checked false statements from partisans are more likely to harness animosity toward political out-groups than fact-checked true statements.

\textbf{Additional robustness checks:} We conduct several checks to validate the robustness of our results: (1) We check for possible multicollinearity issues. The cross-correlations among the independent variables in our analysis are fairly small. Further, the variance inflation factors for the independent variables are all close to one; and, thus, well below the critical threshold of four (see details in Suppl. \nameref{sec:corrs}). Taken together, these findings indicate that multicollinearity is not an issue in our analysis. (2) We exclude fact-checked statements with mixed veracity ratings and repeat our analysis (see details in Suppl. \nameref{sec:no_mixed}). (3) We correct false positives in the manually checked politicized statements and repeat out analysis (see details in Suppl. \nameref{sec:no_false_positives}). (4) We exclude fact-checked statements that contain mentions of both Republicans and Democrats and repeat out analysis (see details in Suppl. \nameref{sec:no_mixed_politicians}). (5) We include additional fixed effects for the fact-checking organizations and specific fact-checkers; see details in Suppl. \nameref{sec:org_fixed}. (6) We include author-level fixed effects, \ie, we additionally control for the party affiliations of the authors of the fact-checked statements (see details in Suppl. \nameref{sec:source_fixed}). (7) We conduct propensity score matching to balance the samples in our dataset and repeat our analysis (see details in Suppl. \nameref{sec:psm}). Our findings are consistently supported by all of these checks.

\section{Discussion}

Politicization plays a critical role in driving engagement with online misinformation \cite{shin_partisan_2017,vosoughi_spread_2018,prollochs_emotions_2021,garrett_conservatives_2021,mosleh_measuring_2022,rathje_out-group_2021,osmundsen_partisan_2021}. As a remedy, fact-checking organizations (\eg, PolitiFact, Snopes) are supposed to identify and debunk politicized misinformation to curb its spread. However, an understanding of how fact-checking platforms pick up politicized misinformation for fact-checking is largely missing. Here, we contribute to research at the intersection of misinformation and politicization by conducting a large-scale empirical analysis characterizing mentions of political elites within true and false statements that have been subject to fact-checking by major \US fact-checking platforms.

\textbf{Implications:}
Our findings imply that major fact-checking organizations in the \US regularly fact-check politicized misinformation, especially during election months. However, there is a pronounced partisan asymmetry: false statements selected for fact-checking are significantly more likely to be pro-Republican, while the selected true statements are more likely to be pro-Democrat. This finding aligns with earlier conjectures that fact-checking organizations in the \US tend to fact-check the political right more than the political left \cite{mena_principles_2019,nieminen_fighting_2019,singh_independent_2024}.
Importantly, however, this observed asymmetry does not necessarily imply political bias on the part of the fact-checking platforms in their reporting of politicized false statements. Instead, it could also reflect the possibility that adherents of the political right produce a higher volume of (viral) misinformation that requires fact-checking. Previous research suggests that widely shared false political statements in the \US are more often pro-Republican \cite{garrett_conservatives_2021,chen_neutral_2021} and that the political right is more likely to hold misconceptions than the political left \cite{mosleh_measuring_2022}. Disentangling these two factors likely requires a representative sample of \textit{all} circulating misinformation. This presents an important -- yet difficult -- challenge for future research.


We further observe a pronounced role of out-group animosity in the politicization of misinformation. Partisans from both sides of the political spectrum frequently harness animosity toward political out-groups in their false statements. This out-group animosity is stronger than usual during election periods. Such a high prevalence of misinformation about political out-groups -- at least within statements that undergo fact-checking -- is worrying as it could imply that more in-group members believe and share this kind of misinformation \cite{rathje_out-group_2021,jenke_affective_2023}.  

From a practical perspective, our study emphasizes the need for close monitoring of the activities and potential biases of major fact-checking organizations such as Snopes and Politifact. These platforms play a significant role in shaping public opinions and discourse in an era of widespread misinformation. Additionally, many research findings depend on the assessments provided by these organizations \cite{vosoughi_spread_2018,allcott_social_2017,prollochs_emotions_2021,garrett_conservatives_2021,rathje_out-group_2021,solovev_moral_2022,mosleh_measuring_2022}. If fact-checking organizations exhibit bias -- whether intentional or unintentional -- it could skew public perception, undermine trust in research findings, and diminish confidence in fact-checking as an (objective) tool to combat misinformation. Hence, it is essential that fact-checking organizations not only conduct their work impartially but also maintain transparency about their practices (\eg, by being open about their criteria for selecting claims). Overall, continuous oversight and transparency are vital for the integrity of the fact-checking process and ensuring that the public (and the academic community) can rely on their evaluations.

\textbf{Limitations and future research:}
Our work has a number of limitations, which can fuel future research as follows. First, our analysis does not reveal whether the observed (political) asymmetries in fact-checking activities reflect political preferences/biases of the fact-checking organizations; or rather a result of adherents to the political left/right producing misinformation at different rates. Yet, disentangling this dual role is challenging as the population of misinformation is unknown, which provides an important challenge for further research. Second, we conduct our analysis based on a large sample of fact-checked statements collected from two fact-checking organizations: Snopes and PolitiFact. These are major fact-checking organizations in the \US and have been used as a main source of fact-checking verdicts in prior misinformation research. To the best of our knowledge, there are currently no other fact-checking sources that can provide comparable fact-checked statements over 15 years. Notwithstanding, to further generalize our findings, one could collect fact-checked statements from other fact-checking sources when available in the future. Third, this study primarily measures politicization through mentions of formal names of political elites, which might miss some other forms of politicization (\eg, nicknames of politicians and non-official political activists). Nevertheless, focusing on mentions of formal names of political elites is a common choice in previous research and ensures that we do not overestimate politicization in false statements \cite{chinn_politicization_2020}. Finally, this paper focuses on statements in English and the context of \US politics. More data from other countries and languages need to be collected to foster a more comprehensive understanding of the link between politicization and misinformation across cultures.

\section{Conclusion}

Political elites are increasingly mentioned in the discussion of social issues and play a critical role in driving engagement with online misinformation. However, an understanding of how fact-checking platforms pick up and report false statements that mention political elites of either Republicans or Democrats remains underexplored. Here, we provide a comprehensive empirical analysis of mentions of political elites in true and false statements that have been subject to fact-checking by two major \US fact-checking organizations. We find that fact-checked false statements are more likely to mention political elites than fact-checked true statements, especially Democrats and during election months. Furthermore, we demonstrate a pronounced role of out-group animosity in politicized misinformation. Specifically, fact-checked false statements that mention political elites carry stronger other-condemning emotions and are more likely to be pro-Republican, while fact-checked true statements are more likely to be pro-Democrat. Taken together, our findings offer new insights into how fact-checking platforms in the \US fact-check politicized misinformation, and bridge important findings on the intersection between misinformation and politicization.

\section{Ethics Statement}
This research has received ethical approval from the Ethics Review Panel of the University of Luxembourg (ref. ERP 23-053 REMEDIS). All analyses are based on publicly available data. We declare no competing interests.


\bibliography{refs}

\clearpage
\begin{center}
\LARGE
\textbf{Supplementary Materials}
\end{center}
\normalsize
\renewcommand\thetable{S\arabic{table}}
\setcounter{table}{0}
\renewcommand\thefigure{S\arabic{figure}}
\setcounter{figure}{0}

\section{Dataset Overview}
\label{sec:data_overview}

An overview of the dataset used in this study is shown in Table \ref{tab:data_summary}.

\section{Manual Validation}
\label{sec:manual_val}

In addition to manually checking 200 random samples, we further use three steps to check the precision of our politician matching approach for all politicized statements:

\emph{Step~1:} We additionally extract the fact-checking comments in the corresponding fact-checking webpages. Fact-checkers provide these comments to explain their verdicts for the veracity of the statements. If the comments also mention other politicians except the politicians mentioned in the original statements. Then the politician names matched in the statements have a high probability of being corresponding politicians. Given this, we match politician names mentioned in the comments. If there is at least one other politician in the comments, we take the associated statements as true positives. As a result, out of the total fact-checked statements that mention politicians, 88.5\% are linked to the comments that mention other politicians. 

\emph{Step~2:} We create a list of specific politicians, including Joe Biden, Kamala Harris, Donald Trump, Barack Obama, Nancy Pelosi, and Hillary Clinton. We take statements that mention these specific names as true positives, because these politician names are somewhat specific to the corresponding politicians in the context of fact-checking. The true positives in this regard are 768 out of \num{1070} statements that are not identified by \emph{Step~1}. 

\emph{Step~3:} We manually check the remaining 302 statements from Steps 1 and 2. We find there are 33 false positives that mention the names same as politician names but do not correspond to the politicians. Finally, we combine the three steps together and find that our politician matching approach has high performance with a precision score of 99.6\%. 

\section{Analysis Across Topics}
\label{sec:analysis_across_topics}

\begin{figure*}[htb]
     \centering
     \begin{subfigure}[b]{0.32\textwidth}
         \centering
         \includegraphics[width=\textwidth]{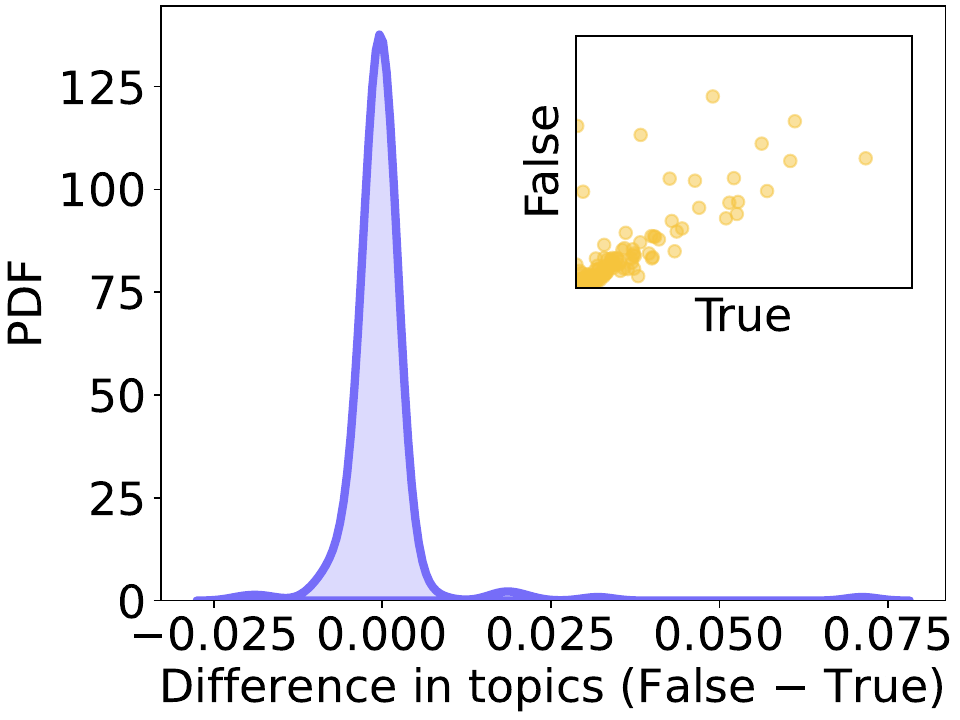}
         \caption{}
         \label{fig:issue_weight_reg_fit}
     \end{subfigure}
     \hfill
     \begin{subfigure}[b]{0.32\textwidth}
         \centering
         \includegraphics[width=\textwidth]{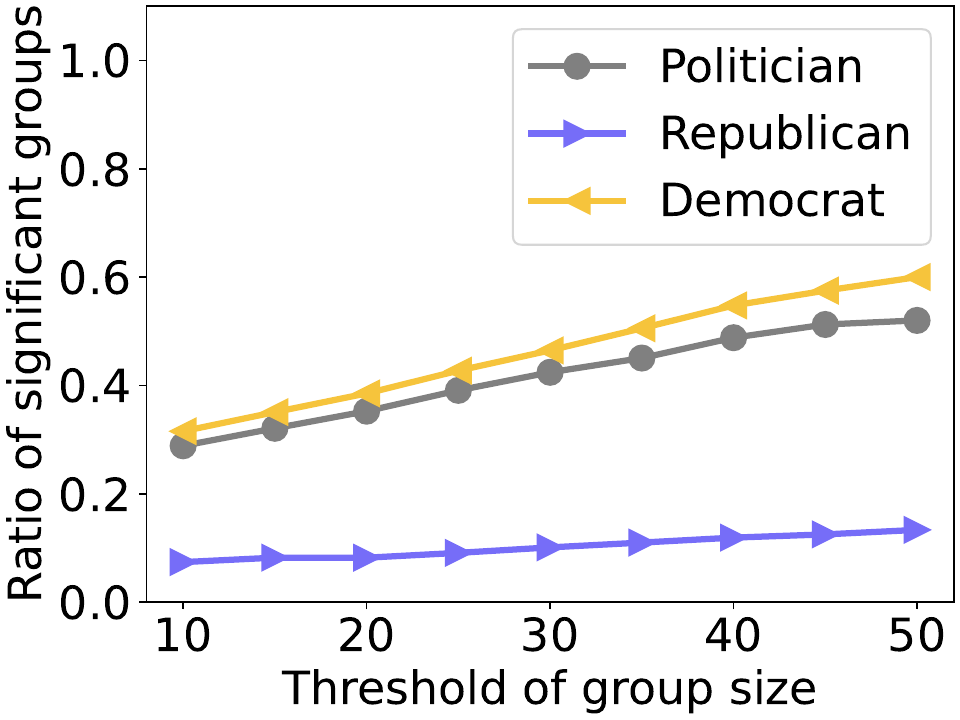}
         \caption{}
         \label{fig:topic_diff_ratio}
     \end{subfigure}
     \hfill
     \begin{subfigure}[b]{0.32\textwidth}
         \centering
         \includegraphics[width=\textwidth]{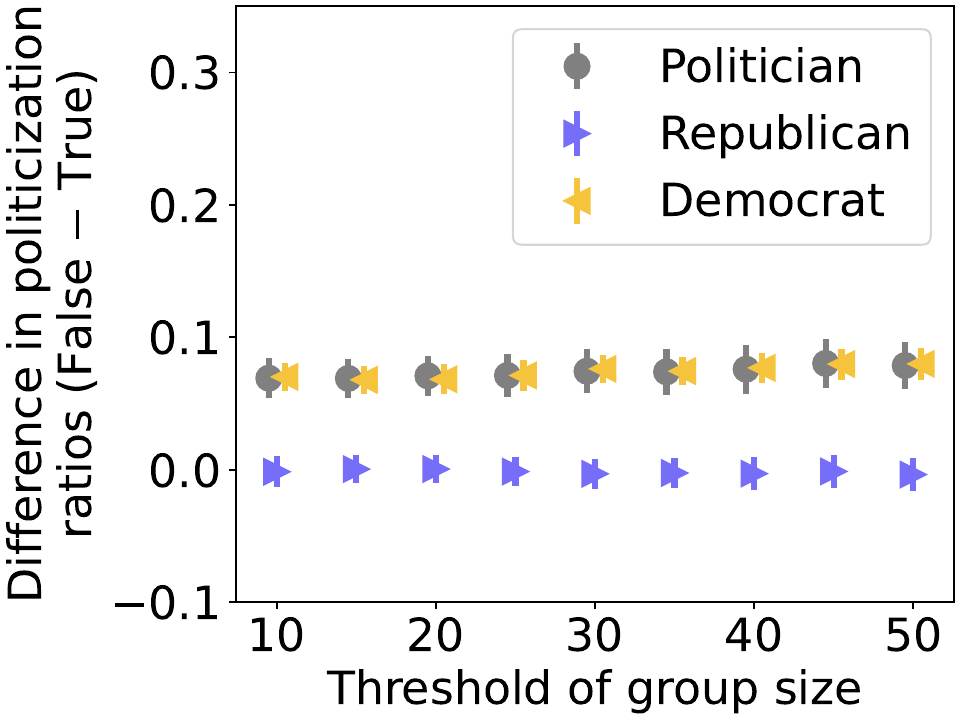}
         \caption{}
         \label{fig:topic_diff_errorbar}
     \end{subfigure}
     \caption{Topics in fact-checked statements. \subref{fig:issue_weight_reg_fit} The distribution of the differences in shared topics between fact-checked true and false statements. The embedded scatters describe the ratios of each shared topic in fact-checked true and false statements. \subref{fig:topic_diff_ratio} The ratios of statement groups in which the mentions of political elites are significantly associated with the veracity of fact-checked statements under different thresholds of group size. \subref{fig:topic_diff_errorbar} The differences in politicization ratios between fact-checked true and false statements under different thresholds of group size. The error bars represent 95\% Confidence Intervals (CIs).}
    \label{fig:topics}
\end{figure*}

The fact-checked statements in our data concern a wide spectrum of 202 topics (see \nameref{sec:materials_and_methods}). Notably, out of the 202 topics, 194 (96\%) are present in both true and false statements. Table \ref{tab:topics} displays the 20 most frequent topics shared by fact-checked true and false statements. Furthermore, as shown in the embedded figure within Fig. \ref{fig:issue_weight_reg_fit}, the ratios of topics in fact-checked true statements have significantly positive correlations with the ratios of corresponding topics in fact-checked false statements ($r=0.751, p<0.001$). Additionally, using a paired two-sample $t$-test ($t=0.009, p>0.05$), we find that the ratios of each shared topic in fact-checked true and false statements do not differ significantly (Fig. \ref{fig:issue_weight_reg_fit}). This suggests that the distribution of statements across topics is not veracity-specific.

While the distribution of topics appears to be independent of veracity (Fig. \ref{fig:issue_weight_reg_fit}), the degree of politicization within fact-checked true and false claims may differ across social issues. We examine how mentions of political elites in fact-checked true and false statements vary across topics. Given that a single statement can discuss multiple topics, we assess each topic within the same statement separately. To this end, we count all fact-checked statements that share the same topic as one statement group. Within each topic-specific statement group, we employ a Chi-Square test to evaluate the association between politicization and veracity. For these in-group tests, it's worth noting that some groups may lack a sufficient number of samples. Therefore, we set a group size threshold ranging from 10 to 50 (in increments of 5) to assess within-group differences. In each Chi-Square test, the input is a 2$\times$2 matrix including the counts of politicized and non-politicized statements in true and false categories.

As shown in Fig. \ref{fig:topic_diff_ratio}, the ratio of statement groups in which mentions of political elites exhibit a statistically significant relationship with veracity relative to all statement groups increases from 28.9\% to 52.0\% with increasing group size thresholds. However, notably, the ratio of statement groups in which mentions of Republicans show a significant association with veracity relative to all statement groups remains relatively low, ranging from 7.4\% to 13.3\%. Conversely, the ratio of statement groups in which mentions of Democrats show a significant relationship with veracity relative to all statement groups rises from 31.5\% to 60.0\%. These findings suggest that across various topics, fact-checked false and true statements consistently show opposing tendencies when it comes to mentioning political elites, particularly Democrats. 

We further consider the politicization ratio within each group and conduct out-group paired $t$-tests between fact-checked true and false statements under each group size threshold. The politicization ratio indicates the ratio of statements mentioning political elites relative to all statements in each group. Fig. \ref{fig:topic_diff_errorbar} shows the variance of the difference in politicization ratio between true and false statements among the topic-specific groups under each threshold for all politicians (grey), Republicans (purple), and Democrats (yellow). We consistently observe that the politicization ratio in fact-checked false statements is significantly higher than that in fact-checked true statements. This disparity primarily arises from the increased likelihood of fact-checked false statements mentioning Democrats, while no statistically significant difference between fact-checked true and false statements is found with regard to mentions of Republicans (Fig. \ref{fig:topic_diff_errorbar}).

Additionally, we keep the group size threshold at 50 and examine the changes in politicization ratios of individual topics between Obama's incumbency and Trump's incumbency. Specifically, 59 topics are presented in both Obama's incumbency and Trump's incumbency, and each has at least 50 fact-checked statements during either Obama's incumbency or Trump's incumbency. There are only nine topics with significantly different politicization ratios between Obama's incumbency and Trump's incumbency (as shown in Fig. \ref{fig:politi_issues_by_incumbency}). The other shared topics have similar politicization ratios during two incumbencies.

\section{Analysis Across Incumbent \US Presidents}
\label{sec:analysis_across_incumbencies}

In the main analysis, we incorporate incumbent fixed effect ($\var{Trump}$) and its interactions with $\var{Falsehood}$ and $\var{Politician}$ to examine the differences in mentions of politicians and other-condemning emotions between Obama's incumbency and Trump's incumbency. The estimation results for the main analysis of mentions of political elites and other-condemning emotions across incumbencies are reported in Table \ref{tab:politician_incumbency_fixed} and Column (1) of Table \ref{tab:emotions_incumbency}, respectively. Subsequently, we further conduct subgroup analysis across incumbent presidents. The estimation results for the mentions of political elites are reported in Table \ref{tab:politician_incumbency}. Columns (1)--(3) of Table \ref{tab:politician_incumbency} report the estimation results for fact-checked statements during Obama's incumbency, and Columns (4)--(6) report the results for fact-checked statements during Trump's incumbency. These results are consistent with our findings in Table \ref{tab:politician_incumbency_fixed}. Additionally, in Table \ref{tab:emotions_incumbency}, the estimation results for the other-condemning emotions in the fact-checked statement subgroups during Obama's [Columns (2)--(4)] and Trump's incumbencies [Columns (5)--(7)] are consistent with the interaction analysis in Column (1).

\section{Political Leanings}
\label{sec:political_leanings}
We first examine the partisan leanings based on all fact-checked statements. The estimation results are reported in Column (1) of Table \ref{tab:partisan_leaning} and used for the main analysis. Additionally, we conduct the same analysis again based on fact-checked statements from PolitiFact. The coefficient estimate of $\var{Falsehood}$ in Column (2) of Table \ref{tab:partisan_leaning} is also significantly positive ($\var{coef.}=0.008, p<0.001$). Within fact-checked statements from PolitiFact, the predicted marginal means for partisan leanings in true and false statements are \num{-0.004} (95\% CI: $[\num{-0.007}, \num{-0.001}]$) and \num{0.004} (95\% CI: $[\num{0.001}, \num{0.006}]$), respectively. The estimates for partisan leanings of fact-checked true and false statements in PolitiFact are consistent with the estimates based on all fact-checked statements.

Subsequently, given that PolitiFact provides author information for the fact-checked statements, we partially identify the political elites who authored the fact-checked statements based on our politician database and incorporate the author information into the regression. The estimation results are reported in Column (3) of Table \ref{tab:partisan_leaning} and used for the main analysis. 

We further examine the moderating effects of $\var{Falsehood}$ and $\var{MTE}$ on the partisan leanings in fact-checked statements from Democrats and Republicans. As shown in Column (4) of Table \ref{tab:partisan_leaning}, the coefficient estimate of $\var{Democrat_{\text{author}} \times Falsehood}$ is significantly negative ($\var{coef.}=-0.014, p<0.01$), and the coefficient estimate of $\var{Republican_{\text{author}} \times Falsehood}$ is significantly positive ($\var{coef.}=0.011, p<0.05$). This indicates that the in-group leanings for Republicans and Democrats are stronger in fact-checked false statements than in fact-checked true statements. Notably, the coefficient estimate of $\var{Falsehood}$ in Column (4) is not statistically significant ($\var{coef.}=0.004, p>0.05$). This suggests that the right-leaning of fact-checked false statements compared to fact-checked true statements is moderated by partisan authors. 

The coefficient estimate for $\var{Democrat_{\text{author}} \times MTE}$ is significantly positive ($\var{coef.}=0.002, p<0.001$), and the coefficient estimate of $\var{Republican_{\text{author}} \times MTE}$ is significantly negative ($\var{coef.}=-0.001, p<0.001$). This suggests the tendency that fact-checked statements from political lefts and rights harness more other-condemning emotions toward political out-groups when elections approach. Finally, we specifically analyze whether fact-checked false statements from Democrats express other-condemning emotions differently compared to fact-checked false statements from Republicans, when mentioning political out-groups. The predicted marginal mean for partisan leaning of fact-checked false statements from Republicans is \num{0.018} (95\% CI: $[\num{0.015}, \num{0.021}]$), while the predicted marginal mean for partisan leaning of fact-checked false statements from Democrats is \num{-0.022} (95\% CI: $[\num{-0.026}, \num{-0.017}]$). The magnitudes of the predicted marginal means in fact-checked false statements from Democrats and Republicans are similar, which indicates that the pattern of harnessing animosity toward political out-groups is shared by both Republicans and Democrats in their fact-checked misinformation stories.

\section{Robustness Checks}

\subsection{Cross-Correlations \& Variance Inflation Factors}
\label{sec:corrs}

The cross-correlations of the independent variables and their variance inflation factors are shown in Table \ref{tab:corr} and Table \ref{tab:vif}, respectively.

\subsection{Statements Without Mixed Veracity}
\label{sec:no_mixed}

Statements with mixed veracity ratings (\eg, half-true and half-false) are included and categorized as false in the main analysis. To ensure the robustness of our findings, we exclude these statements and perform our regression models again with the remaining \num{29596} fact-checked true or false statements. The regression results are reported in Table \ref{tab:politician_no_mixed} and Table \ref{tab:emotions_no_mixed}. We consistently find that fact-checked false statements are more likely to mention political elites, especially Democrats, compared to fact-checked true statements.

\subsection{Statements With Manual Validation}
\label{sec:no_false_positives}

We corrected the false positives identified in Suppl. \nameref{sec:manual_val} and conduct the analysis again. The results remain robust (see Table \ref{tab:politician_manual} and Table \ref{tab:emotions_manual} for details).

\subsection{Statements Without Mixed Political Elites}
\label{sec:no_mixed_politicians}

We exclude statements that mention two or more political elites with mixed party affiliations (\ie, Democrats and Republicans) and conduct the analysis again. The results are robust and consistent with those in the main analysis (see Table \ref{tab:politician_no_mixed_politicians} and Table \ref{tab:emotions_no_mixed_politicians} for details).

\subsection{Organization-Specific Fixed Effects}
\label{sec:org_fixed}

We repeat our analysis with organization-specific fixed effects. The results are reported in Table \ref{tab:politician_org} and Table \ref{tab:emotions_org}. Additionally, we repeat our analysis with fact-checker-specific fixed effects, which allows us to control the heterogeneity of fact-checkers explicitly. The results remain robust and are reported in Table \ref{tab:politician_fact_checker} and Table \ref{tab:emotions_fact_checker}.

\subsection{Author-Specific Fixed Effects}
\label{sec:source_fixed}

We repeat our analysis with author-specific fixed effects to control for the types of authors, \ie, Republicans, Democrats, or others. The results remain robust and are reported in Table \ref{tab:politician_source} and Table \ref{tab:emotions_source}.

\subsection{Propensity Score Matching}
\label{sec:psm}

We conduct propensity score matching for true and false statements to check the robustness of the regression results for RQs1--3. Specifically, we match true and false statements based on the pre-defined variables, \ie, $\var{MTE}$, $\var{Words}$, yearly dummies, organization-specific dummies, and author-specific dummies. With the caliper of 0.001, ``common'' support, and ``noreplacement'' settings, we discard \num{15859} false statements and achieve all biases lower than 0.05 across the independent variables. The regression results based on the matched true and false samples are reported in Table \ref{tab:politician_psm}. Additionally, to validate the results for RQ4, we conduct the same propensity score matching for statements mentioning political elites and statements not mentioning political elites according to the pre-defined variables: $\var{Falsehood}$, $\var{MTE}$, $\var{Words}$, yearly dummies, organization-specific dummies, and author-specific dummies. We achieve all biases lower than 0.05 by discarding 69 statements that mention political elites. We regress the matched statements again for predicting other-condemning emotions and report the results in Table \ref{tab:emotions_psm}. The results are robust and consistent with our findings in the main analysis.

\begin{table*}
\begin{center}
\begin{tabularx}{\textwidth}{@{\hspace{\tabcolsep}\extracolsep{\fill}}l c c c}
\toprule
&{(1)}&{(2)}&{(3)}\\
&{All}&{PolitiFact}&{Snopes}\\
\midrule
Number of statements & {35,014} & {22,561} & {12,453} \\
Observation start date & {01/02/2008} & {01/02/2008} & {01/04/2008} \\
Observation end date & {07/19/2023} & {07/19/2023} & {07/17/2023} \\
Number of fact-checkers & {608} & {585} & {23} \\
Number of topics & {202} & {152} & {66} \\
Politician & {9,308} & {6,363} & {2,945} \\
& {(26.6\%)} & {(28.2\%)} & {(23.6\%)} \\
\quad Republican & {4,403} & {2,881} & {1,522} \\
\quad & {(12.6\%)} & {(12.8\%)} & {(12.2\%)} \\
\quad Democrat & {4,455} & {3,207} & {1,248} \\
& {(12.7\%)} & {(14.2\%)} & {(10.0\%)} \\
Veracity\\
\quad True & {9,616} & {6,199} & {3,417} \\
\quad & {(27.5\%)} & {(27.5\%)} & {(27.4\%)} \\
\quad False & {25,398} & {16,362} & {9,036} \\
\quad & {(72.5\%)} & {(72.5\%)} & {(72.6\%)} \\
Months to elections & {11.603} & {11.078} & {12.555} \\
& {(0.039)} & {(0.048)} & {(0.067)} \\
Other-condemning emotions & {0.096} & {0.092} & {0.103} \\
& {(0.001)} & {(0.001)} & {(0.002)} \\
Words & {18.376} & {18.237} & {18.628} \\
& {(0.042)} & {(0.054)} & {(0.064)} \\
\bottomrule
\end{tabularx}
\caption{Dataset overview. Column (1) reports summary statistics for the whole dataset. Column (2) reports summary statistics for the subset of fact-checked statements from PolitiFact. Column (3) reports summary statistics for the subset of fact-checked statements from Snopes. Percentages or standard errors are included in the parentheses.}
\label{tab:data_summary}
\end{center}
\end{table*}

\begin{table*}
\begin{center}
\begin{tabularx}{\textwidth}{@{\hspace{\tabcolsep}\extracolsep{\fill}}l*{3}{c}}
\toprule
&{(1)}&{(2)}&{(3)}\\
&{True}&{False}&{Total}\\
\midrule
Facebook fact-checks & 326 & 4,009 & 4,335 \\
Politics & 1,315 & 2,663 & 3,978 \\
Fauxtography & 330 & 1,669 & 1,999 \\
Health care & 528 & 1,453 & 1,981 \\
Economy & 699 & 1,130 & 1,829 \\
Elections & 448 & 1,258 & 1,706 \\
Taxes & 517 & 1,107 & 1,624 \\
Coronavirus & 156 & 1,334 & 1,490 \\
Junk news & 3 & 1,412 & 1,415 \\
Crime & 381 & 957 & 1,338 \\
Education & 461 & 845 & 1,306 \\
Immigration & 287 & 935 & 1,222 \\
Public health & 226 & 953 & 1,179 \\
Jobs & 391 & 750 & 1,141 \\
Federal budget & 370 & 743 & 1,113 \\
State budget & 388 & 645 & 1,033 \\
Candidate biography & 297 & 699 & 996 \\
History & 362 & 607 & 969 \\
Fake news & 17 & 839 & 856 \\
Foreign policy & 231 & 584 & 815 \\
\bottomrule
\end{tabularx}
\caption{The twenty most frequent topic tags attached to fact-checked statements. Column (1) reports the number of fact-checked true statements for each topic. Column (2) reports the number of fact-checked false statements for each topic. Column (3) reports the number of all fact-checked statements for each topic.}
\label{tab:topics}
\end{center}
\end{table*}

\begin{table*}
\begin{center}
\begin{tabularx}{\textwidth}{@{\hspace{\tabcolsep}\extracolsep{\fill}}l*{3}{S}}
\toprule
&\multicolumn{1}{c}{(1)}&\multicolumn{1}{c}{(2)}&\multicolumn{1}{c}{(3)}\\
&\multicolumn{1}{c}{Politician}&\multicolumn{1}{c}{Republican}&\multicolumn{1}{c}{Democrat}\\
\midrule
Falsehood &       0.169\sym{**} &      -0.297\sym{***}&       0.634\sym{***}\\
            &     (0.064)         &     (0.077)         &     (0.091)         \\
MTE         &      -0.073\sym{***}&      -0.063\sym{***}&      -0.060\sym{***}\\
            &     (0.004)         &     (0.005)         &     (0.007)         \\
Falsehood $\times$ MTE&       0.028\sym{***}&       0.039\sym{***}&       0.010         \\
            &     (0.005)         &     (0.006)         &     (0.007)         \\
Words       &       0.016\sym{***}&       0.012\sym{***}&       0.013\sym{***}\\
            &     (0.002)         &     (0.002)         &     (0.002)         \\
Trump       &       0.174\sym{*}  &       0.133         &       0.105         \\
            &     (0.069)         &     (0.082)         &     (0.101)         \\
Falsehood $\times$ Trump&      -0.134         &      -0.258\sym{**} &       0.039         \\
            &     (0.072)         &     (0.087)         &     (0.106)         \\
Intercept      &      -1.055\sym{***}&      -1.667\sym{***}&      -2.209\sym{***}\\
            &     (0.080)         &     (0.093)         &     (0.100)         \\
\midrule
Fact-checker REs       &       \checkmark&       \checkmark&       \checkmark\\
\midrule
\#Statements       &       \num{25967}         &       \num{25967}         &       \num{25967}         \\
\bottomrule
\end{tabularx}
\caption{Estimation results for mixed logistic regression models predicting whether the fact-checked statements mention political elites of either political party [Column (1)], Republicans [Column (2)], or Democrats [Column (3)]. Fact-checker-specific random effects and incumbency (Trump) fixed effect are included. Reported are coefficient estimates with standard errors in parentheses. \sym{*} \(p<0.05\), \sym{**} \(p<0.01\), \sym{***} \(p<0.001\).}
\label{tab:politician_incumbency_fixed}
\end{center}
\end{table*}

\begin{table*}
\begin{center}
\begin{tabularx}{\textwidth}{@{\hspace{\tabcolsep}\extracolsep{\fill}}l*{6}{S}}
\toprule
&\multicolumn{3}{c}{Obama}&\multicolumn{3}{c}{Trump}\\
\cmidrule(lr){2-4}\cmidrule(lr){5-7}
&\multicolumn{1}{c}{(1)}&\multicolumn{1}{c}{(2)}&\multicolumn{1}{c}{(3)}&\multicolumn{1}{c}{(4)}&\multicolumn{1}{c}{(5)}&\multicolumn{1}{c}{(6)}\\
&\multicolumn{1}{c}{Politician}&\multicolumn{1}{c}{Republican}&\multicolumn{1}{c}{Democrat}&\multicolumn{1}{c}{Politician}&\multicolumn{1}{c}{Republican}&\multicolumn{1}{c}{Democrat}\\
\midrule
Falsehood &       0.326\sym{***}&      -0.122         &       0.655\sym{***}&      -0.061         &      -0.689\sym{***}&       0.684\sym{***}\\
            &     (0.075)         &     (0.090)         &     (0.104)         &     (0.098)         &     (0.118)         &     (0.143)         \\
MTE         &      -0.088\sym{***}&      -0.075\sym{***}&      -0.072\sym{***}&      -0.051\sym{***}&      -0.045\sym{***}&      -0.040\sym{***}\\
            &     (0.006)         &     (0.007)         &     (0.009)         &     (0.007)         &     (0.008)         &     (0.011)         \\
Falsehood $\times$ MTE&       0.012         &       0.021\sym{*}  &       0.007         &       0.034\sym{***}&       0.049\sym{***}&       0.007         \\
            &     (0.007)         &     (0.008)         &     (0.010)         &     (0.008)         &     (0.009)         &     (0.012)         \\
Words       &       0.007\sym{**} &      -0.004         &       0.015\sym{***}&       0.026\sym{***}&       0.028\sym{***}&       0.011\sym{***}\\
            &     (0.003)         &     (0.003)         &     (0.003)         &     (0.003)         &     (0.003)         &     (0.003)         \\
Intercept      &      -0.697\sym{***}&      -1.252\sym{***}&      -2.136\sym{***}&      -1.311\sym{***}&      -2.011\sym{***}&      -2.290\sym{***}\\
            &     (0.099)         &     (0.112)         &     (0.120)         &     (0.124)         &     (0.142)         &     (0.159)         \\
\midrule
Fact-checker REs       &       \checkmark&       \checkmark&       \checkmark&       \checkmark&       \checkmark&       \checkmark\\
\midrule
\#Statements       &       \num{14515}         &       \num{14515}         &       \num{14515}         &       \num{11452}         &       \num{11452}         &       \num{11452}         \\
\bottomrule
\end{tabularx}
\caption{Estimation results for mixed logistic regression models predicting whether the fact-checked statements mention political elites of either political party [Columns (1) and (4)], Republicans [Columns (2) and (5)], or Democrats [Columns (3) and (6)]. Columns (1)--(3) report results based on the fact-checked statements during Obama's incumbency. Columns (4)--(6) report results based on the fact-checked statements during Trump's incumbency. Fact-checker-specific random effects are included. Reported are coefficient estimates with standard errors in parentheses. \sym{*} \(p<0.05\), \sym{**} \(p<0.01\), \sym{***} \(p<0.001\).}
\label{tab:politician_incumbency}
\end{center}
\end{table*}

\begin{table*}
\begin{center}
\setlength{\tabcolsep}{12pt}
\begin{tabularx}{\textwidth}{@{\hspace{\tabcolsep}\extracolsep{\fill}}l*{7}{S}}
\toprule
&&\multicolumn{3}{c}{Obama}&\multicolumn{3}{c}{Trump}\\
\cmidrule(lr){3-5}\cmidrule(lr){6-8}
&\multicolumn{1}{c}{(1)}&\multicolumn{1}{c}{(2)}&\multicolumn{1}{c}{(3)}&\multicolumn{1}{c}{(4)}&\multicolumn{1}{c}{(5)}&\multicolumn{1}{c}{(6)}&\multicolumn{1}{c}{(7)}\\
\midrule
Politician  &       0.012\sym{***}&       0.011\sym{***}&       0.019\sym{*}  &       0.037\sym{*}  &       0.004         &       0.009         &       0.006         \\
            &     (0.003)         &     (0.003)         &     (0.008)         &     (0.015)         &     (0.004)         &     (0.011)         &     (0.018)         \\
Republican  &                     &                     &                     &      -0.017         &                     &                     &      -0.007         \\
            &                     &                     &                     &     (0.014)         &                     &                     &     (0.016)         \\
Democrat    &                     &                     &                     &      -0.021         &                     &                     &       0.018         \\
            &                     &                     &                     &     (0.014)         &                     &                     &     (0.016)         \\
Falsehood   &       0.007\sym{*}  &       0.006         &       0.005         &       0.005         &       0.012\sym{**} &       0.014\sym{**} &       0.014\sym{**} \\
            &     (0.003)         &     (0.003)         &     (0.004)         &     (0.004)         &     (0.004)         &     (0.005)         &     (0.005)         \\
Politician $\times$ Falsehood&                     &                     &       0.004         &       0.004         &                     &      -0.007         &      -0.012         \\
            &                     &                     &     (0.007)         &     (0.007)         &                     &     (0.010)         &     (0.010)         \\
MTE         &       0.000         &      -0.000         &      -0.000         &      -0.000         &       0.001\sym{*}  &       0.001         &       0.001         \\
            &     (0.000)         &     (0.000)         &     (0.000)         &     (0.000)         &     (0.000)         &     (0.000)         &     (0.000)         \\
Politician $\times$ MTE&                     &                     &      -0.001\sym{**} &      -0.001\sym{**} &                     &       0.000         &       0.000         \\
            &                     &                     &     (0.000)         &     (0.000)         &                     &     (0.001)         &     (0.001)         \\
Words       &       0.001\sym{***}&       0.001\sym{***}&       0.001\sym{***}&       0.001\sym{***}&       0.001\sym{***}&       0.001\sym{***}&       0.001\sym{***}\\
            &     (0.000)         &     (0.000)         &     (0.000)         &     (0.000)         &     (0.000)         &     (0.000)         &     (0.000)         \\
Trump       &       0.014\sym{**} &                     &                     &                     &                     &                     &                     \\
            &     (0.005)         &                     &                     &                     &                     &                     &                     \\
Politician $\times$ Trump&      -0.008         &                     &                     &                     &                     &                     &                     \\
            &     (0.005)         &                     &                     &                     &                     &                     &                     \\
Falsehood $\times$ Trump&       0.005         &                     &                     &                     &                     &                     &                     \\
            &     (0.005)         &                     &                     &                     &                     &                     &                     \\
Intercept      &       0.059\sym{***}&       0.060\sym{***}&       0.056\sym{***}&       0.056\sym{***}&       0.071\sym{***}&       0.070\sym{***}&       0.070\sym{***}\\
            &     (0.004)         &     (0.005)         &     (0.006)         &     (0.006)         &     (0.007)         &     (0.007)         &     (0.007)         \\
\midrule
Fact-checker REs       &       \checkmark&       \checkmark&       \checkmark&       \checkmark&       \checkmark&       \checkmark&       \checkmark\\
\midrule
\#Statements       &       \num{25967}         &       \num{14515}         &       \num{14515}         &       \num{14515}         &       \num{11452}         &       \num{11452}&       \num{11452}\\
\bottomrule
\end{tabularx}
\caption{Estimation results for mixed linear regression models predicting other-condemning emotions in the fact-checked statements. Column (1) reports result for the model that incorporates incumbency (Trump) fixed effect. Columns (2)--(4) report results based on the fact-checked statements during Obama's incumbency. Columns (5)--(7) report results based on the fact-checked statements during Trump's incumbency. Fact-checker-specific random effects are included. Reported are coefficient estimates with standard errors in parentheses. \sym{*} \(p<0.05\), \sym{**} \(p<0.01\), \sym{***} \(p<0.001\).}
\label{tab:emotions_incumbency}
\end{center}
\end{table*}

\begin{table*}
\begin{center}
\begin{tabularx}{\textwidth}{@{\hspace{\tabcolsep}\extracolsep{\fill}}l*{4}{S}}
\toprule
&\multicolumn{1}{c}{(1)}&\multicolumn{1}{c}{(2)}&\multicolumn{1}{c}{(3)}&\multicolumn{1}{c}{(4)}\\
&\multicolumn{1}{c}{All}&\multicolumn{1}{c}{PolitiFact}&\multicolumn{1}{c}{PolitiFact}&\multicolumn{1}{c}{PolitiFact}\\
\midrule
Falsehood &       0.009\sym{***}&       0.008\sym{***}&       0.003\sym{*}  &       0.004         \\
            &     (0.001)         &     (0.002)         &     (0.002)         &     (0.003)         \\
MTE         &      -0.000         &       0.000         &       0.000         &      -0.000         \\
            &     (0.000)         &     (0.000)         &     (0.000)         &     (0.000)         \\
$\var{\text{Democrat}_{\text{author}}}$     &                     &                     &      -0.021\sym{***}&      -0.032\sym{***}\\
            &                     &                     &     (0.002)         &     (0.005)         \\
$\var{\text{Republican}_{\text{author}}}$     &                     &                     &       0.012\sym{***}&       0.013\sym{**} \\
            &                     &                     &     (0.002)         &     (0.005)         \\
$\var{\text{Democrat}_{\text{author}}}$ $\times$ Falsehood&                     &                     &                     &      -0.014\sym{**} \\
            &                     &                     &                     &     (0.004)         \\
$\var{\text{Republican}_{\text{author}}}$ $\times$ Falsehood&                     &                     &                     &       0.011\sym{*}  \\
            &                     &                     &                     &     (0.004)         \\
$\var{\text{Democrat}_{\text{author}}}$ $\times$ MTE&                     &                     &                     &       0.002\sym{***}\\
            &                     &                     &                     &     (0.000)         \\
$\var{\text{Republican}_{\text{author}}}$ $\times$ MTE&                     &                     &                     &      -0.001\sym{***}\\
            &                     &                     &                     &     (0.000)         \\
Words       &       0.000         &       0.000\sym{*}  &       0.000\sym{*}  &       0.000\sym{*}  \\
            &     (0.000)         &     (0.000)         &     (0.000)         &     (0.000)         \\
Intercept      &      -0.038\sym{***}&      -0.049\sym{***}&      -0.042\sym{***}&      -0.040\sym{***}\\
            &     (0.005)         &     (0.006)         &     (0.006)         &     (0.007)         \\
\midrule
Fact-checker REs       &       \checkmark&       \checkmark&       \checkmark&       \checkmark\\
Year FEs     &       \checkmark         &       \checkmark         &       \checkmark    \\ 
\midrule
\#Statements       &       \num{35014}         &       \num{22561}         &       \num{22561}         &       \num{22561}         \\
\bottomrule
\end{tabularx}
\caption{Estimation results for mixed linear regression models predicting partisan leanings in the fact-checked statements. Column (1) reports result based on all fact-checked statements. Columns (2)--(4) report results based on fact-checked statements from PolitiFact. Fact-checker-specific random effects and yearly fixed effects are included. Reported are coefficient estimates with standard errors in parentheses. \sym{*} \(p<0.05\), \sym{**} \(p<0.01\), \sym{***} \(p<0.001\).}
\label{tab:partisan_leaning}
\end{center}
\end{table*}

\begin{table*}
\begin{center}
\begin{tabularx}{\textwidth}{@{\hspace{\tabcolsep}\extracolsep{\fill}}l*{4}{S}}
\toprule
{} & {Politician} & {Falsehood} & {MTE} & {Words} \\
\midrule
Politician & 1.000 &  &  &  \\
Falsehood & 0.050 & 1.000 &  &  \\
MTE & -0.129 & -0.063 & 1.000 &  \\
Words & 0.069 & -0.076 & 0.019 & 1.000 \\
\bottomrule
\end{tabularx}
\caption{Cross-correlations among independent variables.}
\label{tab:corr}
\end{center}
\end{table*}

\begin{table*}
\begin{center}
\begin{tabularx}{\textwidth}{@{\hspace{\tabcolsep}\extracolsep{\fill}}l*{2}{S}}
\toprule
&\multicolumn{1}{c}{VIF}&\multicolumn{1}{c}{1/VIF}\\
\midrule
Politician&1.02&0.976\\
Falsehood&1.01&0.988\\
MTE&1.02&0.980\\
Words&1.01&0.988\\
\bottomrule
\end{tabularx}
\caption{Variance Inflation Factors (VIFs) for the independent variables.}
\label{tab:vif}
\end{center}
\end{table*}

\begin{table*}
\begin{center}
\begin{tabularx}{\textwidth}{@{\hspace{\tabcolsep}\extracolsep{\fill}}l*{3}{S}}
\toprule
            &\multicolumn{1}{c}{(1)}&\multicolumn{1}{c}{(2)}&\multicolumn{1}{c}{(3)}\\
            &\multicolumn{1}{c}{Politician}&\multicolumn{1}{c}{Republican}&\multicolumn{1}{c}{Democrat}\\
\midrule
Falsehood &       0.146\sym{**} &      -0.459\sym{***}&       0.690\sym{***}\\
            &     (0.056)         &     (0.069)         &     (0.078)         \\
MTE         &      -0.041\sym{***}&      -0.033\sym{***}&      -0.042\sym{***}\\
            &     (0.004)         &     (0.005)         &     (0.006)         \\
Falsehood $\times$ MTE&       0.016\sym{***}&       0.026\sym{***}&       0.005         \\
            &     (0.004)         &     (0.005)         &     (0.006)         \\
Words       &       0.024\sym{***}&       0.020\sym{***}&       0.017\sym{***}\\
            &     (0.002)         &     (0.002)         &     (0.002)         \\
Intercept      &      -1.300\sym{***}&      -2.015\sym{***}&      -2.380\sym{***}\\
            &     (0.147)         &     (0.183)         &     (0.177)         \\
\midrule
Fact-checker REs       &       \checkmark&       \checkmark&       \checkmark\\
Year FEs    &       \checkmark         &       \checkmark         &       \checkmark    \\
\midrule
\#Statements       &       \num{29596}         &       \num{29596}         &       \num{29596}         \\
\bottomrule
\end{tabularx}
\caption{Estimation results for mixed logistic regression models predicting whether the fact-checked statements mention political elites of either political party [Column (1)], Republicans [Column (2)], or Democrats [Column (3)]. Fact-checker-specific random effects and yearly fixed effects are included. The statements with mixed veracity are excluded. Reported are coefficient estimates with standard errors in parentheses. \sym{*} \(p<0.05\), \sym{**} \(p<0.01\), \sym{***} \(p<0.001\).}
\label{tab:politician_no_mixed}
\end{center}
\end{table*}

\begin{table*}
\begin{center}
\setlength{\tabcolsep}{12pt}
\begin{tabularx}{\textwidth}{@{\hspace{\tabcolsep}\extracolsep{\fill}}l*{3}{S}}
\toprule
&\multicolumn{1}{c}{(1)}&\multicolumn{1}{c}{(2)}&\multicolumn{1}{c}{(3)}\\
\midrule
Politician  &       0.008\sym{**} &       0.019\sym{**} &       0.023\sym{*}  \\
            &     (0.002)         &     (0.006)         &     (0.011)         \\
Republican  &                     &                     &      -0.004         \\
            &                     &                     &     (0.010)         \\
Democrat    &                     &                     &      -0.004         \\
            &                     &                     &     (0.010)         \\
Falsehood   &       0.011\sym{***}&       0.013\sym{***}&       0.013\sym{***}\\
            &     (0.002)         &     (0.003)         &     (0.003)         \\
Politician $\times$ Falsehood &                     &      -0.008         &      -0.008         \\
            &                     &     (0.005)         &     (0.005)         \\
MTE         &       0.000         &       0.000         &       0.000         \\
            &     (0.000)         &     (0.000)         &     (0.000)         \\
Politician $\times$ MTE   &                     &      -0.001         &      -0.001         \\
            &                     &     (0.000)         &     (0.000)         \\
Words       &       0.001\sym{***}&       0.001\sym{***}&       0.001\sym{***}\\
            &     (0.000)         &     (0.000)         &     (0.000)         \\
Intercept      &       0.100\sym{***}&       0.096\sym{***}&       0.096\sym{***}\\
            &     (0.010)         &     (0.010)         &     (0.010)         \\
\midrule
Fact-checker REs       &       \checkmark&       \checkmark&       \checkmark\\
Year FEs     &       \checkmark         &       \checkmark         &       \checkmark    \\
\midrule
\#Statements       &       \num{29596}         &       \num{29596}         &       \num{29596}         \\
\bottomrule
\end{tabularx}
\caption{Estimation results for mixed linear regression models predicting other-condemning emotions in the fact-checked statements. Fact-checker-specific random effects and yearly fixed effects are included. The statements with mixed veracity are excluded. Reported are coefficient estimates with standard errors in parentheses. \sym{*} \(p<0.05\), \sym{**} \(p<0.01\), \sym{***} \(p<0.001\).}
\label{tab:emotions_no_mixed}
\end{center}
\end{table*}

\begin{table*}
\begin{center}
\begin{tabularx}{\textwidth}{@{\hspace{\tabcolsep}\extracolsep{\fill}}l*{3}{S}}
\toprule
            &\multicolumn{1}{c}{(1)}&\multicolumn{1}{c}{(2)}&\multicolumn{1}{c}{(3)}\\
            &\multicolumn{1}{c}{Politician}&\multicolumn{1}{c}{Republican}&\multicolumn{1}{c}{Democrat}\\
\midrule
Falsehood &       0.181\sym{***}&      -0.308\sym{***}&       0.632\sym{***}\\
            &     (0.052)         &     (0.064)         &     (0.073)         \\
MTE         &      -0.037\sym{***}&      -0.033\sym{***}&      -0.033\sym{***}\\
            &     (0.004)         &     (0.005)         &     (0.006)         \\
Falsehood $\times$ MTE&       0.007         &       0.018\sym{***}&      -0.004         \\
            &     (0.004)         &     (0.005)         &     (0.006)         \\
Words       &       0.023\sym{***}&       0.019\sym{***}&       0.014\sym{***}\\
            &     (0.002)         &     (0.002)         &     (0.002)         \\
Intercept      &      -1.244\sym{***}&      -2.157\sym{***}&      -2.119\sym{***}\\
            &     (0.134)         &     (0.172)         &     (0.153)         \\
\midrule
Fact-checker REs       &       \checkmark&       \checkmark&       \checkmark\\
Year FEs     &       \checkmark         &       \checkmark         &       \checkmark    \\
\midrule
\#Statements       &       \num{35014}         &       \num{35014}         &       \num{35014}         \\
\bottomrule
\end{tabularx}
\caption{Estimation results for mixed logistic regression models predicting whether the fact-checked statements mention political elites of either political party [Column (1)], Republicans [Column (2)], or Democrats [Column (3)]. Mentions of politicians are validated according to Suppl. \nameref{sec:manual_val}. Fact-checker-specific random effects and yearly fixed effects are included. Reported are coefficient estimates with standard errors in parentheses. \sym{*} \(p<0.05\), \sym{**} \(p<0.01\), \sym{***} \(p<0.001\).
}
\label{tab:politician_manual}
\end{center}
\end{table*}

\begin{table*}
\begin{center}
\setlength{\tabcolsep}{12pt}
\begin{tabularx}{\textwidth}{@{\hspace{\tabcolsep}\extracolsep{\fill}}l*{3}{S}}
\toprule
&\multicolumn{1}{c}{(1)}&\multicolumn{1}{c}{(2)}&\multicolumn{1}{c}{(3)}\\
\midrule
Politician  &       0.009\sym{***}&       0.020\sym{***}&       0.022\sym{*}  \\
            &     (0.002)         &     (0.006)         &     (0.010)         \\
Republican  &                     &                     &      -0.003         \\
            &                     &                     &     (0.009)         \\
Democrat    &                     &                     &      -0.002         \\
            &                     &                     &     (0.009)         \\
Falsehood   &       0.007\sym{**} &       0.008\sym{**} &       0.008\sym{**} \\
            &     (0.002)         &     (0.003)         &     (0.003)         \\
Politician $\times$ Falsehood &                     &      -0.003         &      -0.004         \\
            &                     &     (0.005)         &     (0.005)         \\
MTE         &      -0.000         &       0.000         &       0.000         \\
            &     (0.000)         &     (0.000)         &     (0.000)         \\
Politician $\times$ MTE   &                     &      -0.001\sym{*}  &      -0.001\sym{*}  \\
            &                     &     (0.000)         &     (0.000)         \\
Words       &       0.001\sym{***}&       0.001\sym{***}&       0.001\sym{***}\\
            &     (0.000)         &     (0.000)         &     (0.000)         \\
Intercept      &       0.099\sym{***}&       0.095\sym{***}&       0.095\sym{***}\\
            &     (0.009)         &     (0.009)         &     (0.009)         \\
\midrule
Fact-checker REs       &       \checkmark&       \checkmark&       \checkmark\\
Year FEs     &       \checkmark         &       \checkmark         &       \checkmark    \\
\midrule
\#Statements       &       \num{35014}         &       \num{35014}         &       \num{35014}         \\
\bottomrule
\end{tabularx}
\caption{Estimation results for mixed linear regression models predicting other-condemning emotions in the fact-checked statements. Mentions of politicians are validated according to Suppl. \nameref{sec:manual_val}. Fact-checker-specific random effects and yearly fixed effects are included. Reported are coefficient estimates with standard errors in parentheses. \sym{*} \(p<0.05\), \sym{**} \(p<0.01\), \sym{***} \(p<0.001\).
}
\label{tab:emotions_manual}
\end{center}
\end{table*}

\begin{table*}
\begin{center}
\begin{tabularx}{\textwidth}{@{\hspace{\tabcolsep}\extracolsep{\fill}}l*{3}{S}}
\toprule
&\multicolumn{1}{c}{(1)}&\multicolumn{1}{c}{(2)}&\multicolumn{1}{c}{(3)}\\
&\multicolumn{1}{c}{Politician}&\multicolumn{1}{c}{Republican}&\multicolumn{1}{c}{Democrat}\\
\midrule
Falsehood   &       0.192\sym{***}&      -0.308\sym{***}&       0.668\sym{***}\\
            &     (0.054)         &     (0.065)         &     (0.076)         \\
MTE         &      -0.040\sym{***}&      -0.034\sym{***}&      -0.035\sym{***}\\
            &     (0.004)         &     (0.005)         &     (0.006)         \\
Falsehood $\times$ MTE&       0.008         &       0.018\sym{***}&      -0.003         \\
            &     (0.004)         &     (0.005)         &     (0.006)         \\
Words       &       0.020\sym{***}&       0.019\sym{***}&       0.013\sym{***}\\
            &     (0.002)         &     (0.002)         &     (0.002)         \\
Intercept      &      -1.278\sym{***}&      -2.111\sym{***}&      -2.158\sym{***}\\
            &     (0.136)         &     (0.173)         &     (0.158)         \\
\midrule
Fact-checker REs       &       \checkmark&       \checkmark&       \checkmark\\
Year FEs    &       \checkmark         &       \checkmark         &       \checkmark    \\
\midrule
\#Statements       &       \num{34340}         &       \num{34340}         &       \num{34340}         \\
\bottomrule
\end{tabularx}
\caption{Estimation results for mixed logistic regression models predicting whether the fact-checked statements mention political elites of either political party [Column (1)], Republicans [Column (2)], or Democrats [Column (3)]. Fact-checker-specific random effects and yearly fixed effects are included. The statements with mixed political elites are excluded. Reported are coefficient estimates with standard errors in parentheses. \sym{*} \(p<0.05\), \sym{**} \(p<0.01\), \sym{***} \(p<0.001\).}
\label{tab:politician_no_mixed_politicians}
\end{center}
\end{table*}

\begin{table*}
\begin{center}
\setlength{\tabcolsep}{12pt}
\begin{tabularx}{\textwidth}{@{\hspace{\tabcolsep}\extracolsep{\fill}}l*{3}{S}}
\toprule
&\multicolumn{1}{c}{(1)}&\multicolumn{1}{c}{(2)}&\multicolumn{1}{c}{(3)}\\
\midrule
Politician  &       0.009\sym{***}&       0.018\sym{**} &       0.018\sym{**} \\
            &     (0.002)         &     (0.006)         &     (0.006)         \\
Republican  &                     &                     &      -0.001         \\
            &                     &                     &     (0.004)         \\
Falsehood   &       0.007\sym{**} &       0.008\sym{**} &       0.008\sym{**} \\
            &     (0.002)         &     (0.003)         &     (0.003)         \\
Politician $\times$ Falsehood &                     &      -0.002         &      -0.002         \\
            &                     &     (0.005)         &     (0.005)         \\
MTE         &      -0.000         &       0.000         &       0.000         \\
            &     (0.000)         &     (0.000)         &     (0.000)         \\
Politician $\times$ MTE   &                     &      -0.001\sym{*}  &      -0.001\sym{*}  \\
            &                     &     (0.000)         &     (0.000)         \\
Words       &       0.001\sym{***}&       0.001\sym{***}&       0.001\sym{***}\\
            &     (0.000)         &     (0.000)         &     (0.000)         \\
Intercept      &       0.092\sym{***}&       0.089\sym{***}&       0.089\sym{***}\\
            &     (0.009)         &     (0.009)         &     (0.009)         \\
\midrule
Fact-checker REs       &       \checkmark&       \checkmark&       \checkmark\\
Year FEs     &       \checkmark         &       \checkmark         &       \checkmark    \\
\midrule
\#Statements       &       \num{34340}         &       \num{34340}         &       \num{34340}         \\
\bottomrule
\end{tabularx}
\caption{Estimation results for mixed linear regression models predicting other-condemning emotions in the fact-checked statements. Fact-checker-specific random effects and yearly fixed effects are included. The statements with mixed political elites are excluded. Reported are coefficient estimates with standard errors in parentheses. \sym{*} \(p<0.05\), \sym{**} \(p<0.01\), \sym{***} \(p<0.001\).}
\label{tab:emotions_no_mixed_politicians}
\end{center}
\end{table*}

\begin{table*}
\begin{center}
\begin{tabularx}{\textwidth}{@{\hspace{\tabcolsep}\extracolsep{\fill}}l*{3}{S}}
\toprule
            &\multicolumn{1}{c}{(1)}&\multicolumn{1}{c}{(2)}&\multicolumn{1}{c}{(3)}\\
            &\multicolumn{1}{c}{Politician}&\multicolumn{1}{c}{Republican}&\multicolumn{1}{c}{Democrat}\\
\midrule
Falsehood &       0.183\sym{***}&      -0.309\sym{***}&       0.633\sym{***}\\
            &     (0.052)         &     (0.064)         &     (0.073)         \\
MTE         &      -0.037\sym{***}&      -0.033\sym{***}&      -0.032\sym{***}\\
            &     (0.004)         &     (0.005)         &     (0.006)         \\
Falsehood $\times$ MTE&       0.007         &       0.018\sym{***}&      -0.005         \\
            &     (0.004)         &     (0.005)         &     (0.006)         \\
Words       &       0.022\sym{***}&       0.019\sym{***}&       0.015\sym{***}\\
            &     (0.002)         &     (0.002)         &     (0.002)         \\
Intercept      &      -1.213\sym{***}&      -2.163\sym{***}&      -2.078\sym{***}\\
            &     (0.134)         &     (0.172)         &     (0.151)         \\
\midrule
Fact-checker REs       &       \checkmark&       \checkmark&       \checkmark\\
Year FEs     &       \checkmark         &       \checkmark         &       \checkmark    \\
Org FEs     &       \checkmark         &       \checkmark         &       \checkmark    \\ 
\midrule
\#Statements       &       \num{35014}         &       \num{35014}         &       \num{35014}         \\
\bottomrule
\end{tabularx}
\caption{Estimation results for mixed logistic regression models predicting whether the fact-checked statements mention political elites of either political party [Column (1)], Republicans [Column (2)], or Democrats [Column (3)]. Fact-checker-specific random effects, yearly fixed effects, and organization-specific fixed effects are included. Reported are coefficient estimates with standard errors in parentheses. \sym{*} \(p<0.05\), \sym{**} \(p<0.01\), \sym{***} \(p<0.001\).
}
\label{tab:politician_org}
\end{center}
\end{table*}

\begin{table*}
\begin{center}
\setlength{\tabcolsep}{12pt}
\begin{tabularx}{\textwidth}{@{\hspace{\tabcolsep}\extracolsep{\fill}}l*{3}{S}}
\toprule
&\multicolumn{1}{c}{(1)}&\multicolumn{1}{c}{(2)}&\multicolumn{1}{c}{(3)}\\
\midrule
Politician  &       0.009\sym{***}&       0.021\sym{***}&       0.025\sym{*}  \\
            &     (0.002)         &     (0.006)         &     (0.010)         \\
Republican  &                     &                     &      -0.005         \\
            &                     &                     &     (0.009)         \\
Democrat    &                     &                     &      -0.003         \\
            &                     &                     &     (0.009)         \\
Falsehood   &       0.007\sym{**} &       0.008\sym{**} &       0.008\sym{**} \\
            &     (0.002)         &     (0.003)         &     (0.003)         \\
Politician $\times$ Falsehood &                     &      -0.004         &      -0.004         \\
            &                     &     (0.005)         &     (0.005)         \\
MTE         &      -0.000         &       0.000         &       0.000         \\
            &     (0.000)         &     (0.000)         &     (0.000)         \\
Politician $\times$ MTE   &                     &      -0.001\sym{**} &      -0.001\sym{**} \\
            &                     &     (0.000)         &     (0.000)         \\
Words       &       0.001\sym{***}&       0.001\sym{***}&       0.001\sym{***}\\
            &     (0.000)         &     (0.000)         &     (0.000)         \\
Intercept      &       0.098\sym{***}&       0.095\sym{***}&       0.095\sym{***}\\
            &     (0.009)         &     (0.009)         &     (0.009)         \\
\midrule
Fact-Checker REs       &       \checkmark&       \checkmark&       \checkmark\\
Year FEs     &       \checkmark         &       \checkmark         &       \checkmark    \\
Org FEs     &       \checkmark         &       \checkmark         &       \checkmark    \\
\midrule
\#Statements       &       \num{35014}         &       \num{35014}         &       \num{35014}         \\
\bottomrule
\end{tabularx}
\caption{Estimation results for mixed linear regression models predicting other-condemning emotions in the fact-checked statements. Fact-checker-specific random effects, yearly fixed effects, and organization-specific fixed effects are included. Reported are coefficient estimates with standard errors in parentheses. \sym{*} \(p<0.05\), \sym{**} \(p<0.01\), \sym{***} \(p<0.001\).
}
\label{tab:emotions_org}
\end{center}
\end{table*}

\begin{table*}
\begin{center}
\begin{tabularx}{\textwidth}{@{\hspace{\tabcolsep}\extracolsep{\fill}}l*{3}{S}}
\toprule
            &\multicolumn{1}{c}{(1)}&\multicolumn{1}{c}{(2)}&\multicolumn{1}{c}{(3)}\\
            &\multicolumn{1}{c}{Politician}&\multicolumn{1}{c}{Republican}&\multicolumn{1}{c}{Democrat}\\
\midrule
Falsehood &       0.197\sym{***}&      -0.290\sym{***}&       0.641\sym{***}\\
            &     (0.053)         &     (0.065)         &     (0.075)         \\
MTE         &      -0.036\sym{***}&      -0.032\sym{***}&      -0.032\sym{***}\\
            &     (0.004)         &     (0.005)         &     (0.006)         \\
Falsehood $\times$ MTE&       0.006         &       0.017\sym{***}&      -0.005         \\
            &     (0.004)         &     (0.005)         &     (0.006)         \\
Words       &       0.023\sym{***}&       0.019\sym{***}&       0.015\sym{***}\\
            &     (0.002)         &     (0.002)         &     (0.002)         \\
Intercept      &      -1.746\sym{***}&      -2.549\sym{***}&      -2.559\sym{***}\\
            &     (0.445)         &     (0.566)         &     (0.631)         \\
\midrule
Fact-checker FEs     &       \checkmark         &       \checkmark         &       \checkmark    \\
Year FEs    &       \checkmark         &       \checkmark         &       \checkmark    \\
\midrule
\#Statements       &       \num{34323}         &       \num{33692}         &       \num{33870}         \\
\bottomrule
\end{tabularx}
\caption{Estimation results for fixed logistic regression models predicting whether the fact-checked statements mention political elites of either political party [Column (1)], Republicans [Column (2)], or Democrats [Column (3)]. Fact-checker-specific fixed effects and yearly fixed effects are included. Reported are coefficient estimates with standard errors in parentheses. \sym{*} \(p<0.05\), \sym{**} \(p<0.01\), \sym{***} \(p<0.001\).
}
\label{tab:politician_fact_checker}
\end{center}
\end{table*}

\begin{table*}
\begin{center}
\setlength{\tabcolsep}{12pt}
\begin{tabularx}{\textwidth}{@{\hspace{\tabcolsep}\extracolsep{\fill}}l*{3}{S}}
\toprule
&\multicolumn{1}{c}{(1)}&\multicolumn{1}{c}{(2)}&\multicolumn{1}{c}{(3)}\\
\midrule
Politician  &       0.009\sym{***}&       0.020\sym{***}&       0.021\sym{*}  \\
            &     (0.002)         &     (0.006)         &     (0.010)         \\
Republican  &                     &                     &      -0.001         \\
            &                     &                     &     (0.009)         \\
Democrat    &                     &                     &       0.001         \\
            &                     &                     &     (0.009)         \\
Falsehood   &       0.005\sym{*}  &       0.006\sym{*}  &       0.006\sym{*}  \\
            &     (0.002)         &     (0.003)         &     (0.003)         \\
Politician $\times$ Falsehood &                     &      -0.003         &      -0.004         \\
            &                     &     (0.005)         &     (0.005)         \\
MTE         &      -0.000         &       0.000         &       0.000         \\
            &     (0.000)         &     (0.000)         &     (0.000)         \\
Politician $\times$ MTE   &                     &      -0.001\sym{**} &      -0.001\sym{**} \\
            &                     &     (0.000)         &     (0.000)         \\
Words       &       0.001\sym{***}&       0.001\sym{***}&       0.001\sym{***}\\
            &     (0.000)         &     (0.000)         &     (0.000)         \\
Intercept      &       0.119\sym{***}&       0.115\sym{***}&       0.115\sym{***}\\
            &     (0.031)         &     (0.031)         &     (0.031)         \\
\midrule
Fact-checker FEs     &       \checkmark         &       \checkmark         &       \checkmark    \\
Year FEs     &       \checkmark         &       \checkmark         &       \checkmark    \\
\midrule
\#Statements       &       \num{35014}         &       \num{35014}         &       \num{35014}         \\
\bottomrule
\end{tabularx}
\caption{Estimation results for fixed linear regression models predicting other-condemning emotions in the fact-checked statements. Fact-checker-specific fixed effects and yearly fixed effects are included. Reported are coefficient estimates with standard errors in parentheses. \sym{*} \(p<0.05\), \sym{**} \(p<0.01\), \sym{***} \(p<0.001\).
}
\label{tab:emotions_fact_checker}
\end{center}
\end{table*}

\begin{table*}
\begin{center}
\begin{tabularx}{\textwidth}{@{\hspace{\tabcolsep}\extracolsep{\fill}}l*{3}{S}}
\toprule
            &\multicolumn{1}{c}{(1)}&\multicolumn{1}{c}{(2)}&\multicolumn{1}{c}{(3)}\\
            &\multicolumn{1}{c}{Politician}&\multicolumn{1}{c}{Republican}&\multicolumn{1}{c}{Democrat}\\
\midrule
Falsehood &       0.206\sym{***}&      -0.167\sym{*}  &       0.526\sym{***}\\
            &     (0.053)         &     (0.066)         &     (0.074)         \\
MTE         &      -0.037\sym{***}&      -0.031\sym{***}&      -0.036\sym{***}\\
            &     (0.004)         &     (0.005)         &     (0.006)         \\
Falsehood $\times$ MTE&       0.007         &       0.017\sym{***}&      -0.003         \\
            &     (0.004)         &     (0.005)         &     (0.006)         \\
Words       &       0.022\sym{***}&       0.018\sym{***}&       0.015\sym{***}\\
            &     (0.002)         &     (0.002)         &     (0.002)         \\
Intercept      &      -1.455\sym{***}&      -2.617\sym{***}&      -2.117\sym{***}\\
            &     (0.137)         &     (0.184)         &     (0.159)         \\
\midrule
Fact-Checker REs       &       \checkmark&       \checkmark&       \checkmark\\
Year FEs     &       \checkmark         &       \checkmark         &       \checkmark    \\
Author FEs     &       \checkmark         &       \checkmark         &       \checkmark    \\ 
\midrule
\#Statements       &       \num{35014}         &       \num{35014}         &       \num{35014}         \\
\bottomrule
\end{tabularx}
\caption{Estimation results for mixed logistic regression models predicting whether the fact-checked statements mention political elites of either political party [Column (1)], Republicans [Column (2)], or Democrats [Column (3)]. Fact-checker-specific random effects, yearly fixed effects, and author-specific fixed effects are included. Reported are coefficient estimates with standard errors in parentheses. \sym{*} \(p<0.05\), \sym{**} \(p<0.01\), \sym{***} \(p<0.001\).
}
\label{tab:politician_source}
\end{center}
\end{table*}

\begin{table*}
\begin{center}
\setlength{\tabcolsep}{12pt}
\begin{tabularx}{\textwidth}{@{\hspace{\tabcolsep}\extracolsep{\fill}}l*{3}{S}}
\toprule
&\multicolumn{1}{c}{(1)}&\multicolumn{1}{c}{(2)}&\multicolumn{1}{c}{(3)}\\
\midrule
Politician  &       0.010\sym{***}&       0.022\sym{***}&       0.025\sym{*}  \\
            &     (0.002)         &     (0.006)         &     (0.010)         \\
Republican  &                     &                     &      -0.004         \\
            &                     &                     &     (0.009)         \\
Democrat    &                     &                     &      -0.003         \\
            &                     &                     &     (0.009)         \\
Falsehood   &       0.006\sym{*}  &       0.006\sym{*}  &       0.006\sym{*}  \\
            &     (0.002)         &     (0.003)         &     (0.003)         \\
Politician $\times$ Falsehood &                     &      -0.003         &      -0.003         \\
            &                     &     (0.005)         &     (0.005)         \\
MTE         &      -0.000         &       0.000         &       0.000         \\
            &     (0.000)         &     (0.000)         &     (0.000)         \\
Politician $\times$ MTE   &                     &      -0.001\sym{**} &      -0.001\sym{**} \\
            &                     &     (0.000)         &     (0.000)         \\
Words       &       0.001\sym{***}&       0.001\sym{***}&       0.001\sym{***}\\
            &     (0.000)         &     (0.000)         &     (0.000)         \\
Intercept      &       0.109\sym{***}&       0.106\sym{***}&       0.106\sym{***}\\
            &     (0.009)         &     (0.009)         &     (0.009)         \\
\midrule
Fact-Checker REs       &       \checkmark&       \checkmark&       \checkmark\\
Year FEs     &       \checkmark         &       \checkmark         &       \checkmark    \\
Author FEs     &       \checkmark         &       \checkmark         &       \checkmark    \\
\midrule
\#Statements       &       \num{35014}         &       \num{35014}         &       \num{35014}         \\
\bottomrule
\end{tabularx}
\caption{Estimation results for mixed linear regression models predicting other-condemning emotions in the fact-checked statements. Fact-checker-specific random effects, yearly fixed effects, and author-specific fixed effects are included. Reported are coefficient estimates with standard errors in parentheses. \sym{*} \(p<0.05\), \sym{**} \(p<0.01\), \sym{***} \(p<0.001\).
}
\label{tab:emotions_source}
\end{center}
\end{table*}

\begin{table*}
\begin{center}
\begin{tabularx}{\textwidth}{@{\hspace{\tabcolsep}\extracolsep{\fill}}l*{3}{S}}
\toprule
            &\multicolumn{1}{c}{(1)}&\multicolumn{1}{c}{(2)}&\multicolumn{1}{c}{(3)}\\
            &\multicolumn{1}{c}{Politician}&\multicolumn{1}{c}{Republican}&\multicolumn{1}{c}{Democrat}\\
\midrule
Falsehood &       0.317\sym{***}&       0.074         &       0.462\sym{***}\\
            &     (0.063)         &     (0.077)         &     (0.088)         \\
MTE         &      -0.037\sym{***}&      -0.033\sym{***}&      -0.034\sym{***}\\
            &     (0.004)         &     (0.005)         &     (0.006)         \\
Falsehood $\times$ MTE&       0.001         &       0.001         &       0.004         \\
            &     (0.005)         &     (0.006)         &     (0.007)         \\
Words       &       0.022\sym{***}&       0.018\sym{***}&       0.016\sym{***}\\
            &     (0.002)         &     (0.003)         &     (0.003)         \\ 
Intercept      &      -1.566\sym{***}&      -2.818\sym{***}&      -1.975\sym{***}\\
            &     (0.160)         &     (0.208)         &     (0.184)         \\
\midrule
Fact-checker REs       &       \checkmark&       \checkmark&       \checkmark\\
Year FEs     &       \checkmark         &       \checkmark         &       \checkmark    \\
Org FEs     &       \checkmark         &       \checkmark         &       \checkmark    \\
Author FEs     &       \checkmark         &       \checkmark         &       \checkmark    \\
\midrule
\#Statements       &       \num{19155}         &       \num{19155}         &       \num{19155}         \\
\bottomrule
\end{tabularx}
\caption{Estimation results for mixed logistic regression models predicting whether the fact-checked statements mention political elites of either political party [Column (1)], Republicans [Column (2)], or Democrats [Column (3)]. The samples are filtered by propensity score matching. Fact-checker-specific random effects, yearly fixed effects, organization-specific fixed effects, and author-specific fixed effects are included. Reported are coefficient estimates with standard errors in parentheses. \sym{*} \(p<0.05\), \sym{**} \(p<0.01\), \sym{***} \(p<0.001\).
}
\label{tab:politician_psm}
\end{center}
\end{table*}

\begin{table*}
\begin{center}
\setlength{\tabcolsep}{12pt}
\begin{tabularx}{\textwidth}{@{\hspace{\tabcolsep}\extracolsep{\fill}}l*{3}{S}}
\toprule
&\multicolumn{1}{c}{(1)}&\multicolumn{1}{c}{(2)}&\multicolumn{1}{c}{(3)}\\
\midrule
Politician  &       0.009\sym{***}&       0.021\sym{***}&       0.025\sym{*}  \\
            &     (0.002)         &     (0.006)         &     (0.010)         \\
Republican  &                     &                     &      -0.004         \\
            &                     &                     &     (0.009)         \\
Democrat    &                     &                     &      -0.004         \\
            &                     &                     &     (0.009)         \\
Falsehood   &       0.006\sym{*}  &       0.006\sym{*}  &       0.006\sym{*}  \\
            &     (0.002)         &     (0.003)         &     (0.003)         \\
Politician $\times$ Falsehood &                     &      -0.003         &      -0.003         \\
            &                     &     (0.005)         &     (0.005)         \\
MTE         &      -0.000         &       0.000         &       0.000         \\
            &     (0.000)         &     (0.000)         &     (0.000)         \\
Politician $\times$ MTE   &                     &      -0.001\sym{**} &      -0.001\sym{**} \\
            &                     &     (0.000)         &     (0.000)         \\
Words       &       0.001\sym{***}&       0.001\sym{***}&       0.001\sym{***}\\
            &     (0.000)         &     (0.000)         &     (0.000)         \\
Intercept      &       0.110\sym{***}&       0.107\sym{***}&       0.107\sym{***}\\
            &     (0.009)         &     (0.009)         &     (0.009)         \\
\midrule
Fact-checker REs       &       \checkmark&       \checkmark&       \checkmark\\
Year FEs     &       \checkmark         &       \checkmark         &       \checkmark    \\
Org FEs     &       \checkmark         &       \checkmark         &       \checkmark    \\
Author FEs     &       \checkmark         &       \checkmark         &       \checkmark    \\
\midrule
\#Statements       &       \num{34945}         &       \num{34945}         &       \num{34945}         \\
\bottomrule
\end{tabularx}
\caption{Estimation results for mixed linear regression models predicting other-condemning emotions in the fact-checked statements. The samples are filtered by propensity score matching. Fact-checker-specific random effects, yearly fixed effects, organization-specific fixed effects, and author-specific fixed effects are included. Reported are coefficient estimates with standard errors in parentheses. \sym{*} \(p<0.05\), \sym{**} \(p<0.01\), \sym{***} \(p<0.001\).
}
\label{tab:emotions_psm}
\end{center}
\end{table*}

\end{document}